\documentclass[12pt,twoside]{article}

\pdfoutput=1

\usepackage[usenames]{color}
\usepackage{lscape}
\usepackage{amssymb}
\usepackage{rotating}
\usepackage[T1]{fontenc}
\usepackage{graphicx}
\usepackage{fancyhdr}
\usepackage{amsmath}
\usepackage{ae}
\usepackage{aecompl}
\usepackage{floatflt}
\usepackage{hyperref}

\def\pa{\partial}
\def\nn{\nonumber \\}
\def\ov{\overline}

\newlength{\dinwidth}
\newlength{\dinmargin}
\begin{document}

\thispagestyle{empty}
\begin{flushright}
IFT--09--12\\
\end{flushright}

\vspace*{1cm}

\begin{center}
{\Large\bf
Inflation with racetrack superpotential and matter field}
\vspace*{5mm}
\end{center}
\vspace*{5mm} \noindent
\vskip 0.5cm
\centerline{\bf
Marcin Badziak\footnote[1]{mbadziak@fuw.edu.pl}
and Marek Olechowski\footnote[2]{Marek.Olechowski@fuw.edu.pl}
}
\vskip 5mm
\centerline{\em Institute of Theoretical Physics,
University of Warsaw}
\centerline{\em ul.\ Ho\.za 69, PL--00--681 Warsaw, Poland}

\vskip 1cm

\centerline{\bf Abstract}
\vskip 3mm
Several models of inflation with the racetrack superpotential
for the volume modulus coupled to a matter field are investigated.
In particular, it is shown that two classes of racetrack inflation
models, saddle point and inflection point ones, can be constructed
in a fully supersymmetric framework with the matter field $F$-term
as a source of supersymmetry breaking and uplifting.
Two models of $F$-term supersymmetry breaking are considered:
the Polonyi model and the quantum corrected O'Raifeartaigh model.
In the former case, both classes of racetrack inflation models
differ significantly from the corresponding models with non-supersymmetric
uplifting. The main difference is a quite strong dominance of
the inflaton by the matter field. In addition, fine-tuning of the
parameters is relaxed as compared to the original racetrack models.
In the case of the racetrack inflation models coupled to the
O'Raifeartaigh model, the matter field is approximately decoupled from
the inflationary dynamics. In all of the above models the gravitino mass
is larger than the Hubble scale during inflation. The possibility of
having the gravitino much lighter than the Hubble scale is also
investigated. It is very hard to construct models with light gravitino
in which the volume modulus dominates inflation. On the other hand,
models in which the inflationary dynamics is dominated by the matter
field are relatively simple and seem to be more natural.

\newpage

\section{Introduction}

What kind of fundamental physics stands behind inflation in the very
early Universe? When looking for the answer to the above question one
often points towards the string theory. The intensive studies of inflation
in the string theory began with the invention of stable de Sitter (dS)
vacua in the famous KKLT paper \cite{kklt}. Construction of such vacua
is inevitable for inflationary model building because the inflaton (the
field which drives inflation) should end its evolution in this kind
of vacuum. The KKLT procedure consists of three steps. In the first
step, the dilaton and the complex structure moduli are stabilized in
a supersymmetry (SUSY) preserving minimum by turning on some non-trivial
fluxes \cite{Giddings}. In the second step, the K\"ahler moduli, including
the volume modulus, are stabilized by some non-perturbative effects such
as gaugino condensation or instantons.\footnote{For the recent discussion on the reliability of the effective theory for the K\"ahler moduli see \cite{addG,Gallego,Brizi}. }
At this stage SUSY remains
unbroken and the vacuum energy is negative. In the last step, the
$\ov{D3}$-branes are introduced to break supersymmetry and to uplift
the minimum to a dS space. However, in the effective field theoretical
description the $\ov{D3}$-branes break supersymmetry explicitly.
This is the main drawback of the KKLT model. The last step of the KKLT
mechanism has been improved in the papers
\cite{matter}-\cite{duplifting_lust} in which dS vacua were obtained
due to spontaneous SUSY breaking.

Many models of inflation have been constructed within the KKLT
framework \cite{stringinf}. Especially interesting scenarios are those
in which the volume modulus plays the role of the inflaton.
Such models using a racetrack superpotential for the modulus,
involving two non-perturbative terms, were proposed in
\cite{racetrack,lw}. With this kind of superpotential two different
inflationary scenarios have been realized.
In the first of them, inflation takes place in the vicinity of a
saddle point of the potential with the axion, associated with the
volume modulus, being the inflaton \cite{racetrack}. In the second
scenario, the real part of the volume modulus is the inflaton and
inflation takes place in the vicinity of an inflection point of the
potential \cite{lw}. Unfortunately, in both scenarios SUSY breaking
and uplifting are due to $\ov{D3}$-branes which break supersymmetry
explicitly.

Even though many forms of manifestly supersymmetric uplifting have been
developed, a vast majority of them have not been applied to inflationary
scenarios. To the best of our knowledge only two papers address the issue
of spontaneous SUSY breaking in racetrack inflation models. The $D$-term
uplifting \cite{achucarro} of the saddle point racetrack inflation model
\cite{racetrack} was performed in \cite{dupracetrack}. The string
theory $\alpha'$-corrections were used to realize saddle point racetrack
inflation in a fully supersymmetric framework in \cite{westphal}. In the
latter approach the volume modulus $F$-term is responsible for uplifting
and SUSY breaking.

In the present work we investigate $F$-term uplifting
in both, saddle point \cite{racetrack} and inflection point \cite{lw},
racetrack inflation scenarios. We assume that the source of SUSY breaking
is a hidden sector matter field.
We consider two models of $F$-term SUSY breaking. The first one is the
Polonyi model \cite{polonyi}, while the second one is the quantum
corrected O'Raifeartaigh model \cite{O'Raifeartaigh,okklt}. In the case of the racetrack
inflation model coupled to the Polonyi model, we show that both
scenarios (saddle point and inflection point ones)
significantly differ from the original ones with
$\ov{D3}$-brane uplifting. In our scenarios, the matter field dominates
SUSY breaking as well as inflationary dynamics. In addition, fine-tuning
of the parameters required for successful inflation is relaxed as
compared to original models with non-SUSY uplifting, especially in the
inflection point scenario. The situation is much different in the case
of models coupled to the quantum corrected O'Raifeartaigh model. We find,
rather unexpectedly, that the matter field responsible for SUSY breaking
can be decoupled from the inflationary dynamics. This is a realization
of ``pure'' $F$-term uplifting which does not alter significantly
any feature of the original model except the fact that supersymmetry
is broken spontaneously.

In most of the inflationary models based on the KKLT moduli stabilization,
including those described above, the gravitino mass in the post-inflationary
vacuum exceeds the Hubble scale during inflation \cite{kl}.
Typically, the scale of
inflation is much larger than the electroweak scale while the
gravitino mass sets a lower bound on the scale of SUSY breaking. This
implies that such models are incompatible with a TeV-scale supersymmetry.
The above mentioned correlation between the scale of inflation and the
gravitino mass can be avoided if inflation ends in a SUSY (near)
Minkowski minimum \cite{kl}. However, it is not easy to construct a model
realizing this idea. It was found in \cite{bo,brustein,covi2} that the
slow-roll parameter $\eta$ is smaller than $-2/3$ for tree-level
K\"ahler potential and arbitrary superpotential so the slow-roll
condition $|\eta|\ll1$ is violated.
This problem can be cured by adding string corrections to the K\"ahler
potential. With such corrections a triple gaugino condensation model
was constructed which accommodates arbitrary small gravitino mass and
a high scale of inflation \cite{bo}. It was later shown \cite{bo2}
that the number of the gaugino condensates can be reduced to two,
as in the standard racetrack model, by admitting (effectively) positive
exponents in the non-perturbative terms in the superpotential.\footnote{The problem of disentangling the gravitino mass from the Hubble scale during inflation was investigated also in \cite{lvgrav} (in the context of Large Volume Scenario) and  \cite{openracetrack} (in the context of D-brane inflation).}

A part of the present work is devoted to the investigation of the impact
of a matter field on models of volume modulus inflation ending in
a SUSY (near) Minkowski minimum. In particular, we study a model with
the racetrack superpotential with two positive exponents coupled to
a matter field. The main issue we address is whether, and under what
circumstances, inflation can be realized without inclusion of any
string corrections to the K\"ahler potential. It occurs that the matter
field always plays a dynamical role in inflation and cannot be decoupled.

We propose a model of inflection point inflation in which inflation is totally dominated by the matter
field. The advantage of this model is that it can lead to inflation
for any combination of the signs of the exponents in the racetrack
superpotential. Matter field inflationary models have been constructed
before in \cite{matterturz,deCarlos}. However, our model is qualitatively
different because it can accommodate an arbitrarily light gravitino.

The model that we propose resembles the one constructed many years ago in \cite{Holman}. However, the model of \cite{Holman} involves only the matter field so it is not coupled to the moduli sector. It is
very important to study such coupling since the moduli fields are
always present in supergravity models if the string theory is assumed
to be the underlying theory. The impact of the moduli sector on the
well established supersymmetric inflationary models such as chaotic
\cite{chaoticsugra} or hybrid inflation \cite{hybrid} was studied in
\cite{Brax1,Brax2,Brax3,Davis1,Davis2}. In typical situations the
presence of the moduli spoils inflation or at least significantly
constrains the parameter space. Therefore, it is very encouraging
that the inflection point model can be
naturally coupled to the volume modulus.

The paper is organized as follows.
In section \ref{condsec} general constraints on the stability of dS
vacua and slow-roll inflation originating from the K\"ahler potential
are recalled. Section \ref{racesec} contains a brief review of
inflationary models based on the racetrack superpotential.
Inflation in such models coupled to a matter field sector is
investigated in section \ref{racemattersec}. Two kinds of scenarios
are considered. In the first, the Polonyi field $F$-term
is used to uplift the racetrack potential. In the second,
the uplifting is realized by the quantum corrected O'Raifeartaigh model.
In section \ref{scaleSUSYsec} the impact of the matter field on
inflationary models admitting low scale of SUSY breaking is analysed.
Section \ref{conclsec} contains the summary of our results.

\section{Conditions for K\"ahler potential}
\label{condsec}

In this section we recall the conditions which a supergravity model
has to satisfy in order to allow for stable dS vacua and/or slow-roll
inflation. The F-term contribution to the supergravity
potential is given by
\begin{equation}
        \label{potentialG}
        V=e^G\left(G^{i\ov{j}}G_i G_{\ov{j}}-3\right) \,,
\end{equation}
with $G$ being the standard combination
of the superpotential $W$ and the K\"ahler potential $K$:
\begin{equation}
        \label{G}
        G(\Phi_i,\Phi_i^{\dag})=K(\Phi_i,\Phi_i^{\dag})
+\log W(\Phi_i)+\log \ov{W}(\Phi_i^{\dag})
\,.
\end{equation}
As usually, subscripts on $G$ denote appropriate derivatives,
$G_i\equiv\frac{\pa G}{\pa\Phi_i}$,
$G_{\ov{i}}\equiv\frac{\pa G}{\pa\Phi_i^\dag}$, while
$G^{i\ov{j}}$ is the matrix inverse to the matrix of the second
derivatives $G_{i\ov{j}}$.

Conditions for the existence of stable non-supersymmetric Minkowski
vacua were found in \cite{Grs,Grs2,Grs3} and generalized to
de Sitter vacua in \cite{covi}. It was pointed out in \cite{bo}
that similar analysis can be performed to constrain models of
slow-roll inflation. In that work, conditions for viable inflation
were found. Those results were then generalized to arbitrary number
of fields in \cite{covi2}. There are similarities between analysis
of de Sitter (dS) vacua and slow-roll inflation. Stable dS vacuum
requires by definition that the potential has positive curvature
in all directions. The situation is very similar in the case
of slow-roll inflation with the only difference that the inflaton
direction is almost flat but tachyonic. Bearing in mind this small
difference, in the following we present conditions for both stable
dS vacua and slow-roll inflation. The key quantity in this discussion
is the curvature tensor $R_{i\ov{j}p\ov{q}}$ of the K\"ahler manifold
which is defined by the metric given by the second derivative of the
K\"ahler potential $G_{i\ov{j}}$. The necessary condition for stable
dS vacua (or slow-roll inflation) is given by \cite{covi,covi2}:
\begin{equation}
\label{etacon}
R(f^i)<\frac{2}{\widehat{G}^2} \,,
\end{equation}
where $R(f^i)\equiv R_{i\ov{j}p\ov{q}}f^if^{\ov{j}}f^pf^{\ov{q}}$
is the sectional curvature along the direction of SUSY
breaking (i.e.\ the Goldstino direction) and
$f_i\equiv G_i/\widehat{G}$ is the unit vector defining that
direction. We also introduced the quantity
$\widehat{G}\equiv\sqrt{G^iG_i}$ related in a simple way
to the value of the potential: $\widehat{G}^2=3+e^{-G}V$.
Notice that the Minkowski vacuum corresponds to $\widehat{G}^2=3$.
In \cite{covi2} the upper bound on the slow-roll parameter $\eta$
(which is defined as the lowest eigenvalue of the $\eta$-matrix)
was found in terms of the sectional curvature along the Goldstino
direction\footnote{
We use a notation which differs from that used in \cite{covi2},
where quantity $\gamma\equiv V/(3e^G)$ was introduced.
These notations are related by $\widehat{G}^2=3(1+\gamma)$.
}:
\begin{equation}
\label{etamax}
\eta\leq\eta_{\rm max}\equiv
-\frac{2}{3}+\frac{4}{\widehat{G}}\sqrt{\epsilon}+
\frac{\widehat{G}^2-3}{\widehat{G}^2}\epsilon
+\frac{\widehat{G}^2}{\widehat{G}^2-3}\left(\frac{2}{3}-R(f^i)\right)
\,,
\end{equation}
where $\epsilon$ is the second slow-roll parameter which should be
very small during inflation. The condition (\ref{etacon}) (as well
as (\ref{etamax})) in its full generality is very complicated.
However, it can be a source of many useful informations for some
specific
K\"ahler geometries.
First of all, in a model with only
one field, say $T$, condition (\ref{etacon}) reduces to:
\begin{equation}
\label{RTcon}
R_T<\frac{2}{\widehat{G}^2}
\,,
\end{equation}
where $R_T$ is the curvature scalar of the K\"ahler manifold.
For $T$ being the volume modulus with the K\"ahler potential:
\begin{equation}
\label{kahler}
K=-3\ln(T+\ov{T}) \,,
\end{equation}
the curvature scalar equals $R_T=2/3$. Therefore, condition
(\ref{RTcon}) implies that for the above K\"ahler potential and
arbitrary superpotential neither a stable Minkowski or dS vacuum
nor slow-roll inflation can be constructed. This shows how powerful
the constraints on the curvature of the K\"ahler manifold can be.
One can show that in models with the K\"ahler potential
(\ref{kahler}) the trace of the $\eta$-matrix
equals $-4/3$ at any non-SUSY stationary point \cite{bo}.
Therefore, the parameter $\eta$ is always smaller than $-2/3$,
as can be seen from (\ref{etamax}). Since condition (\ref{RTcon})
does not depend on the superpotential, the only way to overcome above
difficulties is to modify the K\"ahler potential. Restricting to the
one field only, one can add subleading corrections to (\ref{kahler})
such as $\alpha'$ \cite{alphaprime} or string loop corrections
\cite{stringloop}. Inflationary models incorporating such corrections
were constructed in \cite{westphal,bo,bo2}.

Instead of invoking correction to the K\"ahler potential
one may consider additional fields in the theory.
In this paper we analyze models in which inflation and
supersymmetry breaking are driven by two fields,
the volume modulus and a matter field $\Phi$.
Whether the addition of a matter field can help in fulfilling
the necessary condition (\ref{etacon}), depends on the structure
of the K\"ahler potential. One way of coupling a matter
field to the modulus is to use the following no-scale K\"ahler
potential:
\begin{equation}
\label{kahlernoscale}
K=-3\ln(T+\ov{T}-|\Phi|^2) \,.
\end{equation}
However, in this case the K\"ahler manifold is a maximally symmetric
coset space with a constant curvature $R(f^i)=2/3$ \cite{Grs2}.
Therefore, the matter field does not change the curvature and
slow-roll inflation cannot be
realized\footnote{
In fact, if the D-terms contribute to SUSY breaking, condition
(\ref{etacon}) can be somewhat relaxed \cite{Grs3}. In \cite{dupracetrack}
such situation was explicitly realized by uplifting racetrack inflation
model \cite{racetrack} using D-terms and K\"ahler potential of the form
(\ref{kahlernoscale}). In the present paper we neglect contribution
from the D-term potential.
}.
Another class of models for which condition (\ref{etacon}) is relatively
simple is that for which the K\"ahler potential is separable
into a sum of terms, each depending only on a single field.
In such a case the necessary condition for slow-roll inflation reduces to:
\begin{equation}
\label{thetacon}
\sum_i
R_i\Theta_i^4<\frac{2}{\widehat{G}^2} \,,
\end{equation}
where $R_i$ are the scalar curvatures of the one dimensional
submanifolds associated with each of the fields and
$\Theta_i^2\equiv G_{i\ov{i}}f^if^{\ov{i}}$
(no summation over $i$ or $\ov{i}$) are the spherical coordinates
parameterizing SUSY breaking \cite{Grs}.
They satisfy the condition $\sum_i\Theta_i^2=1$ following from
the fact that $f^i$ is a unit vector. A simple example of a model with
separable K\"ahler potential is the one with the volume modulus coupled
to the canonically normalized matter field:
\begin{equation}
\label{kahlermatter}
K=-3\ln(T+\ov{T})+|\Phi|^2 \,.
\end{equation}
For a canonically normalized field, the scalar curvature $R_{\Phi}$
vanishes, so condition (\ref{thetacon}) can be rewritten as:
\begin{equation}
\label{thetaTcon}
\Theta_T^4<\frac{3}{\widehat{G}^2} \,.
\end{equation}
So, it is clear that the matter field contributing to SUSY breaking
can help in constructing a stable dS vacuum and slow-roll inflation.
In the limiting case $\Theta_T\rightarrow0$, when the matter field
very strongly dominates SUSY breaking, the necessary condition
(\ref{thetaTcon}) is satisfied independently of the value
of $\widehat{G}^2$.

\section{Volume modulus inflation with racetrack superpotential}
\label{racesec}

In this section we describe two models of inflation driven by the
volume modulus $T=t+i\tau$ which were proposed in \cite{racetrack}
and \cite{lw}. Both models have the same basic structure.
The superpotential has the racetrack form:
\begin{equation}
\label{Wrace}
W^{(T)}=A+Ce^{-cT}+De^{-dT} \,,
\end{equation}
and the K\"ahler potential is given by the tree level formula
(\ref{kahler}). A non-supersymmetric
uplifting term:
\begin{equation}
\label{uplift}
\Delta V=\frac{E}{t^2} \,,
\end{equation}
originating from $\ov{D3}$-branes is used to obtain the correct
value of the vacuum energy. Even though both models have the same
elements, different parts of the parameter space are explored
in each case leading to significantly different inflationary
scenarios. Therefore, we discuss these models separately in the
following subsections.

\subsection{Saddle point inflation}

\begin{figure}[t!]
  \centering
  \includegraphics[width=9cm,height=6cm,angle=0]{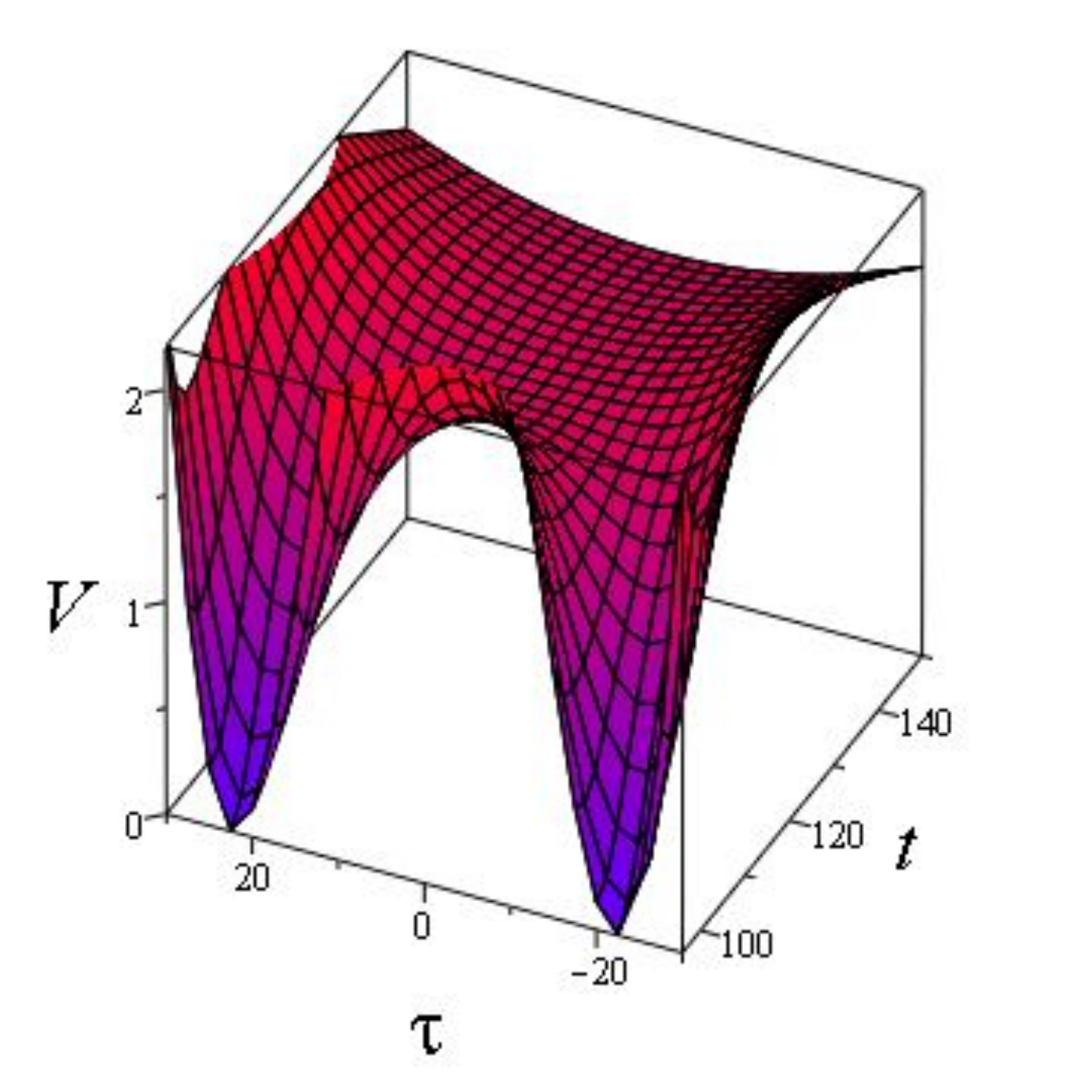}
  \caption{Inflationary part of the potential for the saddle point
model of \cite{racetrack} with non-SUSY uplifting (\ref{uplift}).}
  \label{racetrack3d}
\end{figure}

The first model of volume modulus inflation was proposed in
\cite{racetrack}. In that model inflation takes place in the
vicinity of a saddle point of the potential
(located at $\tau=0$). The inflaton, which is
mainly the axion $\tau$, rolls down slowly towards one of the
minima at $\tau\neq0$. The shape of the inflationary part of
the potential is shown in figure \ref{racetrack3d}.
The potential is $Z_2$-symmetric in the inflaton direction
$\tau\leftrightarrow-\tau$. This leads to a generic upper bound
for the spectral index $n_s\lesssim0.95$, as was pointed out in
\cite{lw,nsbrax}. The key ingredient of that model is the uplifting
term (\ref{uplift}). An obvious reason to include such a term is
the resulting possibility to adjust the cosmological constant in
the post-inflationary vacuum. However, the positive contribution
to the cosmological constant is not the only role of the
uplifting term (\ref{uplift}). It is also crucial for the stability
of the vacuum and the possibility of long enough slow-roll inflation
for the following reason: The uplifting term (\ref{uplift}) explicitly
breaks supersymmetry, so the constraint (\ref{RTcon}) on the K\"ahler
geometry is no longer valid. Without this term the model would be
fully supersymmetric so for the K\"ahler potential (\ref{kahler})
neither stable dS (or Minkowski) vacuum nor slow-roll inflation
could be constructed (as was discussed in the previous section).
This model requires also some fine-tuning of the
parameters and of the initial conditions. The flux parameter $A$
must be adjusted with the precision of order $10^{-4}$.

\subsection{Inflection point inflation}

\begin{figure}[t]
\centering
  \includegraphics[width=8cm,angle=0]{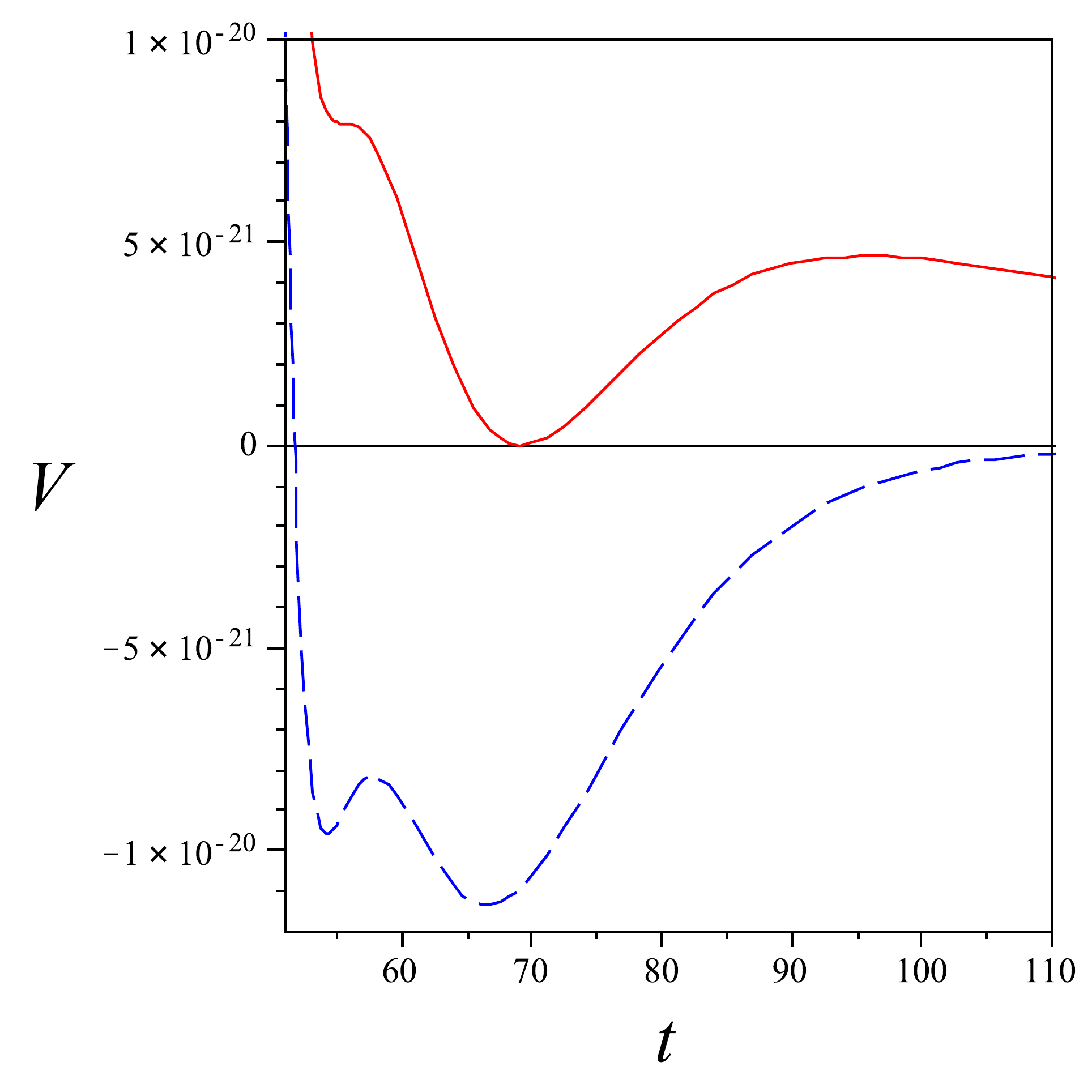}
  \caption{The potential  for $\tau=0$  before (dashed line)
and after (solid line) uplifting by the non-SUSY term (\ref{uplift})
in the model of inflection point inflation \cite{lw}.
}
\label{plottKL}
\end{figure}

It was pointed out in \cite{lw} that within models with the racetrack
superpotential (\ref{Wrace}) also inflection point inflation can be
realized. In such scenario the field $t$ plays the role of the inflaton
while $\tau=0$ during and after inflation. The construction of that model
relies on the fact that for some region of the parameter space the
potential without uplifting has two AdS local minima, as shown in
figure \ref{plottKL}. The minimum at larger value of the volume $t$
is deeper than the one at smaller volume. The uplifting term (\ref{uplift})
is used to lift both minima to dS spaces.
The value of $E$ is chosen in such a way that the minimum at larger
volume has positive but almost vanishing cosmological constant.
The minimum at smaller volume is more strongly lifted and acquires
a large cosmological constant. It is lifted also more strongly than
the local maximum separating both minima. So, for strong enough lifting,
this local maximum and the minimum at smaller volume may disappear.
With appropriate tuning of the parameters one can obtain a very flat
inflection point where inflation can take
place\footnote{
Alternatively, with slightly different tuning of the parameters
one can obtain a very flat saddle point. In such a case the spectral
index is bounded from above
$n_s\lesssim0.93$ \cite{lw}
}.
The spectral index predicted by that model depends on the number
of e-folds of inflation $N$ \cite{Baumann2,Sanchez,mssm,lw} and lies
in the range $0.93\lesssim n_s\lesssim1$. The smallest value
$n_s\approx0.93$ can be obtained in the limit of very large $N$.

Characteristic feature of that model is the relation  between the
fine-tuning (of parameters and initial conditions) and the height
of the barrier separating the vacuum from the run-away region \cite{bo2}.
Namely, more fine-tuning of the parameters (and also of initial
conditions) is required for a higher barrier. It was found
in \cite{bo2} that the inflaton does not overshoot the barrier if the
fine-tuning of parameters (e.g.\ $A$) is at least at the level
of $10^{-8}$. This fine-tuning is much stronger than in the
saddle point inflation model recalled in the previous subsection.

In both models described in this section, the uplifting term
(\ref{uplift}) is crucial for obtaining both a stable dS or Minkowski
vacuum and a region suitable for slow-roll inflation.
The main drawback of both models is that this uplifting term
explicitly breaks supersymmetry. One of the main goals of the present
work is to construct fully supersymmetric versions of the above
two models. Instead of the explicitly non-supersymmetric uplifting
term (\ref{uplift}), a hidden sector matter field will be used as
a source of uplifting and SUSY breaking.

\section{Racetrack inflation with uplifting from matter fields}
\label{racemattersec}

The possibility of constructing dS vacua with the help of hidden
sector matter fields was suggested first in \cite{matter}.
In this section we investigate viability of inflationary
models based on the racetrack superpotential (\ref{Wrace})
with supersymmetric uplifting provided by a matter field.
We investigate models for which the superpotential and the K\"ahler
potential are sums of terms depending on the volume modulus, $T$,
and the matter field, $\Phi$:
\begin{equation}
W=W^{(T)}(T)+W^{(\Phi)}(\Phi)
\,,
\qquad\qquad
K=K^{(T)}(T,\ov{T})+K^{(\Phi)}(\Phi,\ov{\Phi})
\,,
\end{equation}
where $W^{(T)}$ and $K^{(T)}$ are given by (\ref{Wrace})
and (\ref{kahler}), respectively.
As we will show later, it is enough to consider the $\Phi$-dependent
part of the superpotential having a simple linear form
\begin{equation}
\label{Wmatter}
W^{(\Phi)}=c_0-\mu^2\Phi
\,.
\end{equation}
In the following subsections we investigate two different forms
of $K^{(\Phi)}$. For both of them inflationary models can be
constructed but the dynamics of inflation differs significantly.

\subsection{Polonyi uplifting}

We start with the canonical K\"ahler potential for the matter field:
\begin{equation}
\label{kahlercan}
K^{(\Phi)}=\Phi\ov{\Phi}
\,.
\end{equation}
The model with such K\"ahler potential and the superpotential
(\ref{Wmatter}) is the well-known Polonyi model of
supersymmetry breaking \cite{polonyi}. In that model one can find,
for appropriate choice of the parameters, a SUSY breaking Minkowski
minimum at
\begin{equation}
\phi_{\rm min}=\sqrt{3}-1 , \hspace{1cm} \theta_{\rm min}=0
\,,
\label{Polonyi_min}
\end{equation}
where we expressed the complex matter field in terms of its real
components as $\Phi=\phi+i\theta$. One can expect that after coupling
the Polonyi model to the modulus sector, the vev of $\Phi$ will be not
much different from that given in (\ref{Polonyi_min}).
This was explicitly verified in \cite{Abe,Lebedev} where moduli
stabilization in the KKLT model coupled to the Polonyi model was studied.
The moduli stabilization is a local problem in a sense that the
only issue which matters is the stability of the potential at
a Minkowski (or dS) stationary point.
On the other hand, the problem of constructing
an inflationary model involves also the global structure of the
potential. The reason is that the Minkowski vacuum and the inflationary
region are in separate domains of the field space. A priori one
cannot be sure that there always exists a trajectory connecting
these two regions. It is especially not obvious that such a
trajectory exists when one increases the dimensionality of the
field space by introducing a matter field. Therefore, the possibility
of successful moduli stabilization with the non-SUSY uplifting
substituted by the matter uplifting does not immediately imply
that also inflation can be realized with this new form of uplifting.
Nevertheless, we found
that it is possible to construct both inflationary models based on
the racetrack superpotential coupled to the Polonyi model. In what
follows we present examples of such models.

\subsubsection{Inflection point inflation with Polonyi uplifting}
\label{section_InfPol}

\begin{figure}
  \centering
  \includegraphics[width=8cm,angle=0]{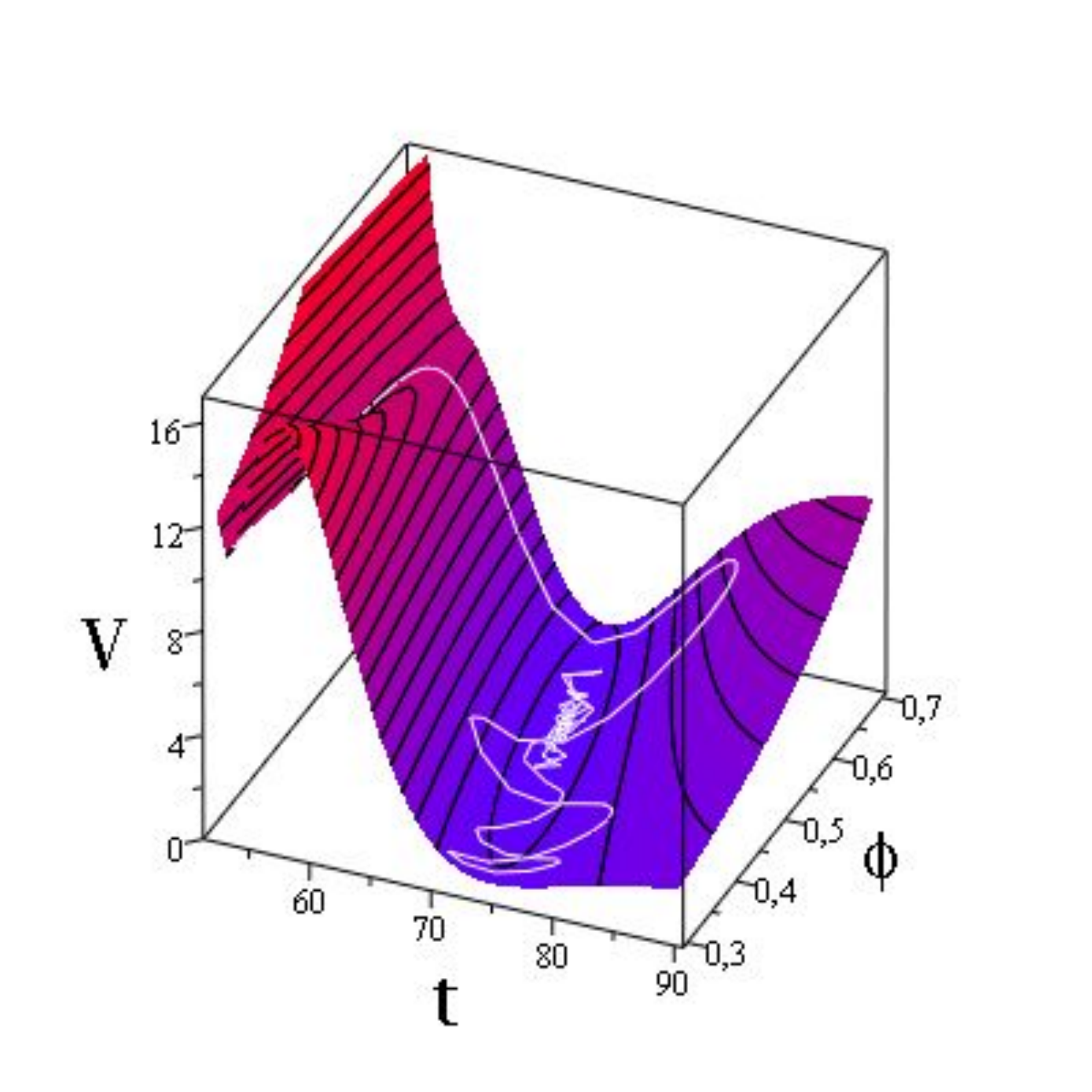}
  \includegraphics[width=8cm,angle=0]{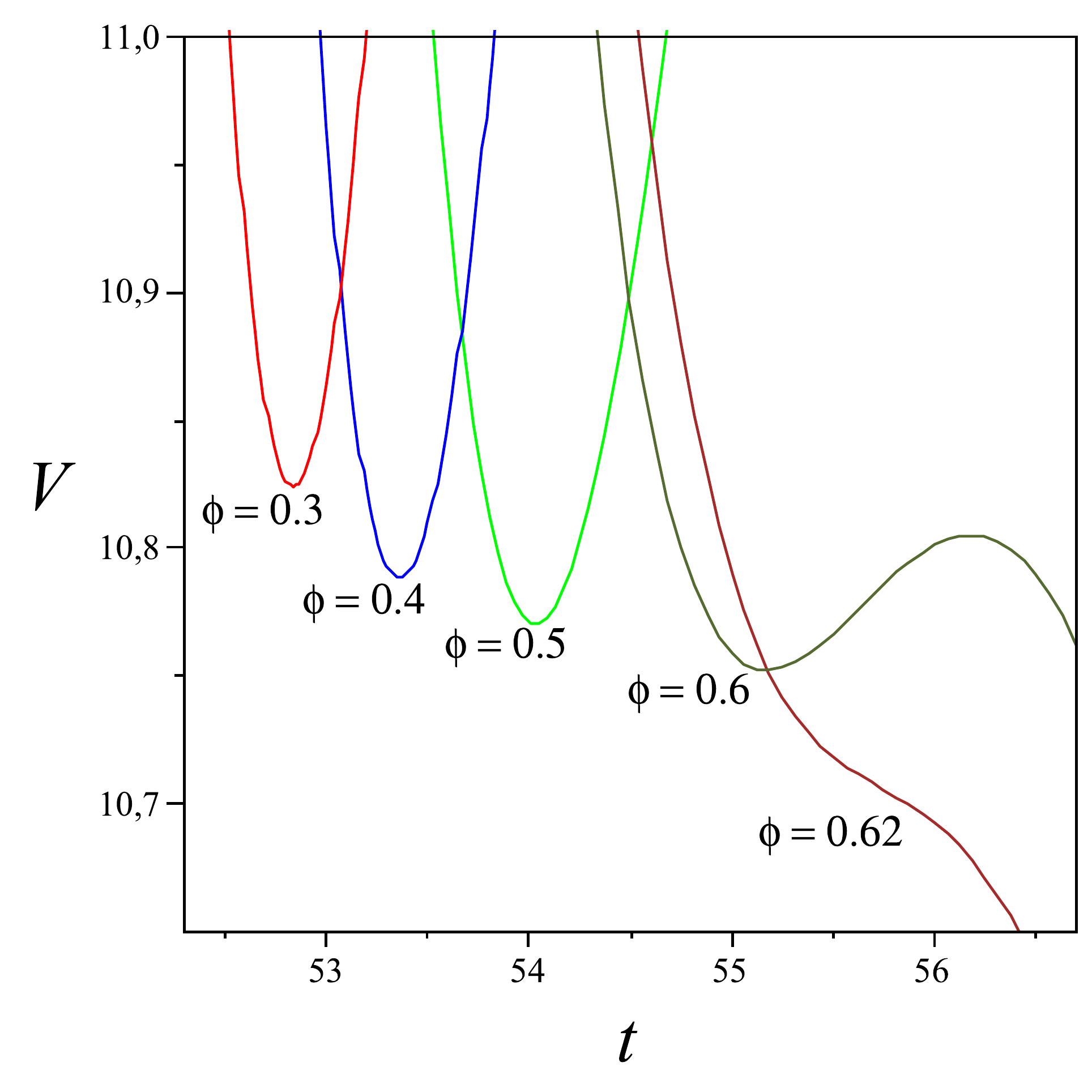}
  \caption{
Left panel: The potential multiplied by $10^{15}$ for the inflection
point model coupled to the Polonyi sector with the parameters
(\ref{parKL_polonyi}) and $\tau=\theta=0$. The white curve represents
the field trajectory from the inflationary inflection point to the
Minkowski vacuum. Right panel: Several $\phi=$const.\ sections
of the inflationary valley.
}
  \label{KL_polonyi}
\end{figure}

We apply the Polonyi uplifting first to the inflection point model
\cite{lw}. This case is easier to investigate because the imaginary
parts of the fields are settled at $\tau=\theta=0$ during and after
inflation and the inflationary dynamics is described effectively by
only two real fields $t$ and $\phi$ (all four real fields will be
involved in the saddle point inflation described in the next subsection).
To illustrate the main features of the model we consider an
example with the following set of parameters:
\begin{equation}
\begin{tabular}{lll}
$W_0=-5.393\cdot10^{-5} , \qquad$ &
$C=-\frac{2}{5} \,, \qquad$ &
$D=\frac{2}{50} \,,$
\\[4pt]
$\mu^2=1.097814\cdot10^{-4} \, , \qquad$ &
$c=\frac{2\pi}{40} \, , \qquad$ &
$d=\frac{2\pi}{60} \, ,$
\end{tabular}
\label{parKL_polonyi}
\end{equation}
where $W_0\equiv A+c_0$ is the sum of the field-independent terms
generated by fluxes and present in $W^{(T)}$ and $W^{(\Phi)}$.
The value of $\mu^2$ is chosen to adjust the cosmological constant
in the post-inflationary vacuum. The potential in the ($t$-$\phi$)-plane
is shown in figure \ref{KL_polonyi}. Inflation takes place in the
vicinity of a very flat inflection point located
at $t_{\rm inf}\approx54.18$, $\phi_{\rm inf}\approx0.5295$.
At this inflection point the previously introduced quantities
related to the cosmological constant and the SUSY breaking
have the following values: $\widehat{G}^2|_{\rm inf}\approx6.4$,
$\Theta_{\Phi}^2|_{\rm inf}\approx0.987$ (we consider only two
complex fields so of course
$\Theta_{T}^2|_{\rm inf}=1-\Theta_{\Phi}^2|_{\rm inf}$).
The value of $\Theta_{\Phi}^2|_{\rm inf}$ so close to unity indicates
that supersymmetry breaking during inflation is strongly dominated
by the matter field. So, one can expect that the sectional curvature
along the direction of SUSY breaking should have a value close to
that of the curvature of the field subspace spanned by $\Phi$, which
vanishes in the case of the considered model.
Numerical calculations confirm this expectation giving
a very small number $R(f^i)|_{\rm inf}\approx0.00012$. Substituting
the obtained values of $R(f^i)|_{\rm inf}$ and $\widehat{G}^2|_{\rm inf}$
into (\ref{etamax}), one finds the following upper bound on the $\eta$
parameter: $\eta < \eta_{\rm max} \approx0.59$. This maximal value of
$\eta$ is positive so condition (\ref{etamax}) is not an obstacle in
realizing  slow-roll inflation. Indeed, there are 3 positive eigenvalues
of the $\eta$-matrix at the inflection point: $\eta_2\approx0.87$,
$\eta_3\approx223$, $\eta_4\approx334$ (of course $\eta_1$ vanishes
at the inflection point). Moreover, the second slow-roll parameter
is small enough to allow for inflation: $\epsilon\approx3\cdot10^{-9}$.

Equations of motion describing the evolution of the fields are collected
in the Appendix. Using (\ref{eqmot}) we numerically evolved the fields
initially set at the inflection point with vanishing velocities.
The obtained evolution corresponds to about $111$ e-folds of inflation.
It can be seen from
figure \ref{KL_polonyi} that inflation proceeds along the valley
(mainly) in the $\phi$-direction. After violation of the slow-roll
conditions at $t_{\rm end}\approx56.71$, $\phi_{\rm end}\approx0.635$,
the inflaton rolls down (first mainly in $t$-direction which becomes
unstable at $\phi\approx0.62$) towards Minkowski vacuum located at
$t_{\rm vac}\approx70.07$, $\phi_{\rm vac}\approx0.5619$,
where the evolution ends after a period of oscillations.
Supersymmetry breaking at the Minkowski vacuum, similarly as
near the inflection point, is dominated by the matter field
($\Theta_{\Phi}^2|_{\rm vac}\approx0.966$). Thus, the necessary condition
(\ref{etacon}) for the stability of the Minkowski vacuum is easily
satisfied.

We computed also the CMB signatures for this model (appropriate
formulae can be found in the Appendix). The parameters (\ref{parKL_polonyi})
were chosen in such a way as to generate the density perturbations
with the amplitude $\frac{\delta\rho}{\rho}\approx2\cdot 10^{-5}$
consistent with COBE measurements (this will be the case also for all
other examples presented in this paper). We found the spectral index
$n_s\approx0.95$ at the COBE normalization scale corresponding
to about $55$ e-folds before the end of inflation.
This value is consistent with the 5-year WMAP result
$n_s=0.96\pm0.013$ \cite{wmap}. We checked that for parameters other than
(\ref{parKL_polonyi}) the value
of the spectral index can be approximated, with the precision
better than one percent, by the analytic formula:
\begin{equation}
\label{fns}
n_s\approx1
-\frac{2\pi}{N_{\rm tot}}\cot\left(\frac{\pi N_e}{2N_{\rm tot}}\right)
\,,
\end{equation}
where $N_{\rm tot}$ is the total number of e-folds during inflation
starting exactly at the inflection point while $N_e$ is the number
of e-folds between the time when a given scale crosses the horizon
and the end of inflation. The above formula was derived in
\cite{Baumann2,Sanchez,mssm,lw} for one-field inflection point
inflation. It occurs that inclusion of the matter field does not alter
significantly the prediction for the spectral index.
We calculated also the tensor to scalar perturbations ratio,
$r\approx10^{-7}$, and the running of the spectral index,
$\frac{d n_s}{d\ln k}\approx10^{-3}$. These values are negligible
from the observational point of view.

The important lesson which follows from this example is that it is
possible to realize inflection point inflation without
non-SUSY uplifting (\ref{uplift}). However, some
features of the model with SUSY broken by the Polonyi field are
substantially different from that of the original model \cite{lw}.
First of all, in the present model the volume modulus no longer plays
the role of the inflaton. The main component of the tachyonic
mass eigenstate during inflation is the field $\phi$
(at the inflection point this mass eigenstate has the components
($t$, $\phi$, $\tau$, $\theta$)=(0.1438, 0.9896, 0, 0)).
The mass squared eigenvalues during the period of slow-roll
inflation are:
$m^2_{\phi_{\rm inf}}\approx -(4\cdot10^{10} {\rm GeV})^2$,
$m^2_{\theta_{\rm inf}}\approx(2\cdot10^{11} {\rm GeV})^2$,
$m^2_{t_{\rm inf}}\approx(4\cdot10^{12} {\rm GeV})^2$,
$m^2_{\tau_{\rm inf}}\approx(5\cdot10^{12} {\rm GeV})^2$,
where the subscripts denote the main component of a given
mass eigenstate. One can see that the volume modulus, $t$, and
the axion associated with it, $\tau$, are at least one order of
magnitude heavier than the imaginary component of the matter field,
$\theta$. The pattern of masses at the Minkowski vacuum is similar:
$m^2_{\phi_{\rm vac}}\approx(3\cdot10^{11} {\rm GeV})^2$,
$m^2_{\theta_{\rm vac}}\approx(10^{11} {\rm GeV})^2$,
$m^2_{t_{\rm vac}}\approx(2\cdot10^{12} {\rm GeV})^2$,
$m^2_{\tau_{\rm vac}}\approx(2\cdot10^{12} {\rm GeV})^2$
(of course there is no any tachyonic state at the vacuum).
The gravitino at the Minkowski vacuum is very heavy:
$m_{3/2}\approx2\cdot10^{11} {\rm GeV}$.
Such a large gravitino mass is a characteristic feature of models
with a deep AdS minimum before uplifting. In such models the gravitino
mass is related to the scale of inflation and typically exceeds the
value of the Hubble constant during inflation,  as was pointed out
in \cite{kl}. In our example $H\approx10^{11} {\rm GeV}$ so it is
indeed smaller than the gravitino mass. Notice that all non-tachyonic
scalars have during inflation masses bigger than the Hubble scale $H$,
so in the model with the parameters (\ref{parKL_polonyi})
no substantial isocurvature perturbations are generated.

\begin{table}[t]
\centering
\begin{tabular}{|c|c|c|c|c|c|c|}
\hline	
$c$ & $\Delta$ & $m_{\theta_{\rm inf}}$[GeV] & $H$[GeV] & $m_{3/2}$[GeV] & $\widehat{G}^2$ & $\eta_2$
\\ \hline
$\frac{2\pi}{57}$ & $10^{-8}$ & $3\cdot10^{9}$  & $4\cdot10^{8}$  & $2\cdot10^{9}$  &  $3.1$ & $14.06$
\\ \hline
$\frac{2\pi}{55}$ & $7\cdot10^{-8}$ & $4\cdot10^{9}$  & $9\cdot10^{8}$  & $3\cdot10^{9}$  &  $3.2$ & $7.36$
\\ \hline
$\frac{2\pi}{50}$ & $10^{-6}$ & $10^{10}$  & $5\cdot10^{9}$  & $9\cdot10^{9}$  &  $3.7$ & $3.02$
\\ \hline
$\frac{2\pi}{45}$ & $8\cdot10^{-6}$ & $5\cdot10^{10}$  & $2\cdot10^{10}$  & $3\cdot10^{10}$  &  $4.4$ & $1.67$
\\ \hline
$\frac{2\pi}{40}$ & $9\cdot10^{-5}$ & $2\cdot10^{11}$  & $10^{11}$  & $2\cdot10^{11}$  &  $6.4$ & $0.89$
\\ \hline
$>\frac{2\pi}{40}$ & \multicolumn{6}{|c|}{no inflection point at $\phi>0$} \\ \hline
\end{tabular}
\caption{Dependence of various quantities on the parameter $c$.
The parameter $\mu^2$  is always adjusted to keep vanishing cosmological
constant at the minimum while $W_0$ is always fine-tuned in order to
have a very flat inflection point appropriate for inflation.
Other parameters are chosen as in (\ref{parKL_polonyi}) up to
a constant which multiplies all  of the superpotential parameters
in order to satisfy COBE normalization.  }
\label{tab_KL_polonyi}
\end{table}
\begin{table}[t]
\centering
\begin{tabular}{|c|c|c|c|c|c|c|}
\hline	
$c$ & $\Delta$ & $m_{\theta_{\rm inf}}$[GeV] & $H$[GeV] & $m_{3/2}$[GeV] & $\widehat{G}^2$ & $\eta_2$
\\ \hline
$<\frac{2\pi}{40}$ & \multicolumn{6}{|c|}{barrier separates inflection point from Minkowski vacuum} \\
\hline
$\frac{2\pi}{40}$ & $2\cdot10^{-3}$ & $8\cdot10^{11}$  & $2\cdot10^{12}$  & $2\cdot10^{12}$  &  $47.6$ & $0.09$
\\ \hline
$\frac{2\pi}{35}$ & $10^{-3}$ & $6\cdot10^{11}$  & $10^{12}$  & $10^{12}$  &  $31.4$ & $0.14$
\\ \hline
$\frac{2\pi}{30}$ & $6\cdot10^{-4}$ & $4\cdot10^{11}$  & $5\cdot10^{11}$  & $5\cdot10^{11}$  &  $22.3$ & $0.19$
\\ \hline
$>\frac{2\pi}{30}$ & \multicolumn{6}{|c|}{overshooting problem} \\ \hline
\end{tabular}
\caption{The same as in table \ref{tab_KL_polonyi} for inflation starting at the inflection point with $\phi<0$. }
\label{tab_KL_polonyi_negphi}
\end{table}

\begin{figure}[t]
  \centering
  \includegraphics[width=8cm,angle=0]{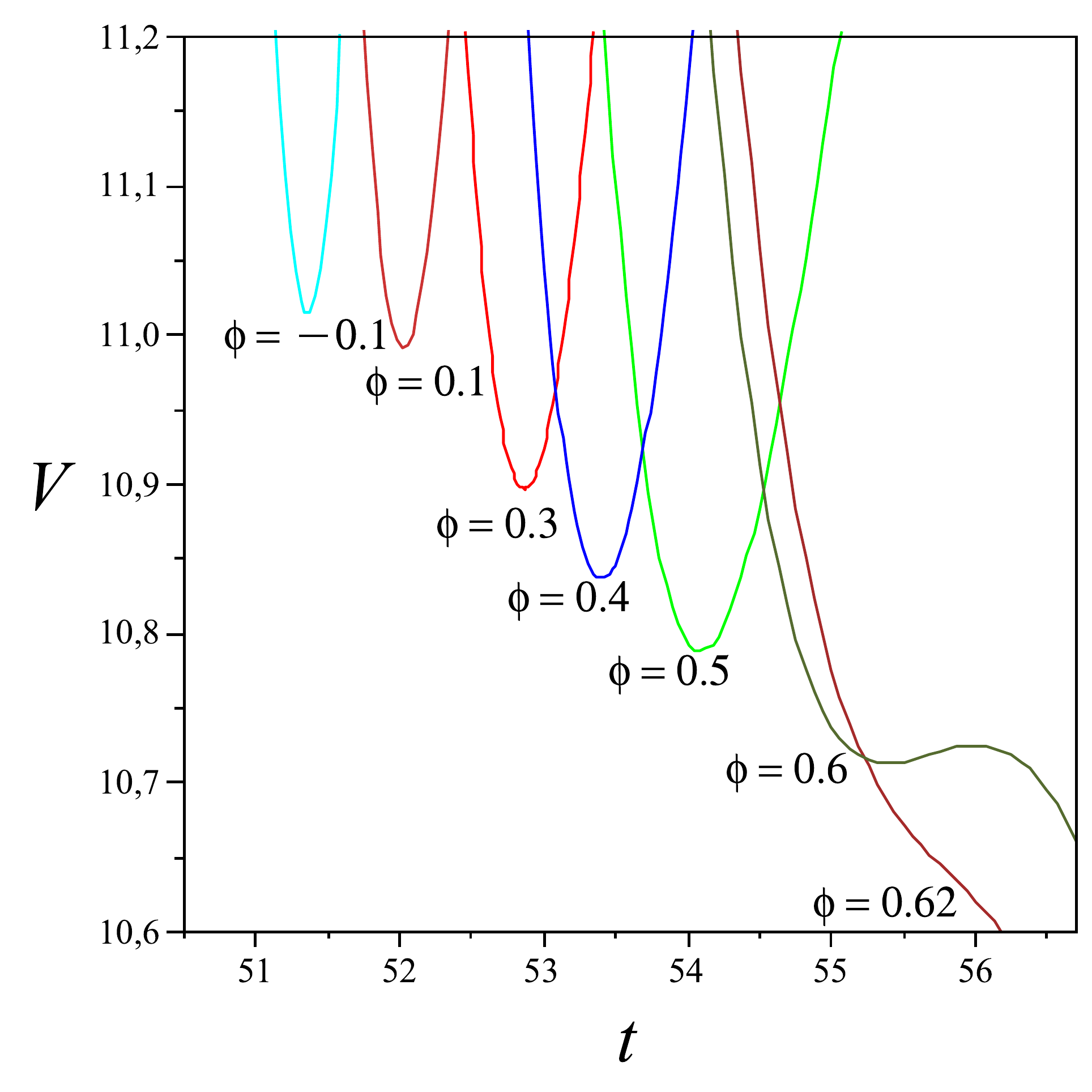}
  \caption{
Sections of the inflationary valley for inflection point model
coupled to the Polonyi sector with inflationary inflection point
situated at $\phi<0$. The parameters are chosen as in (\ref{parKL_polonyi})
except $W_0=-5.43\cdot10^{-5}$ which is adjusted to ensure the flatness of
the inflection point at $\phi_{\rm inf}\approx-0.103$,
$t_{\rm inf}\approx51.35$ (the parameter $\mu^2$ is chosen to cancel the
cosmological constant in the post-inflationary vacuum).
Notice that the
value of $W_0$ differs only by about one percent from that required for
a very flat inflection point at $\phi>0$. This is the reason why the
inflationary valley for $\phi>0$ is flat enough to prolong inflation.
}
  \label{KL_polonyi_negphi}
\end{figure}

The second important difference of the present model as compared
to the model \cite{lw} with non-SUSY uplifting is that the fine-tuning
of the parameters is no longer strictly related to the overshooting
problem. In order to quantify this fine-tuning, we introduce the
following quantity:
\begin{equation}
\Delta_{\rm }\equiv\left|\frac{W_{0_{\rm min}}-W_{0_{\rm max}}}
{\frac{1}{2}(W_{0_{\rm min}}+W_{0_{\rm max}})}\right| \,,
\end{equation}
where $W_{0_{\rm min}}$ ($W_{0_{\rm max}}$) is the minimal (maximal) value
of the parameter $W_0$ for which one can obtain at least $60$ e-folds of
inflation and the spectral index in the range $0.92<n_s<1$
(i.e.\ consistent at 3$\sigma$ with the 5-year WMAP result \cite{wmap})
while keeping other parameters fixed (except $\mu^2$ which is always
adjusted to cancel the cosmological constant in the vacuum).
Less fine-tuned models correspond to higher values of $\Delta_{\rm }$.
For the model with the parameters (\ref{parKL_polonyi}) $\Delta_{\rm }$
is of the order $10^{-4}$. Thus, the range of the parameters which allow
for inflation is 4 orders of magnitude larger than the corresponding
range in a model with non-SUSY uplifting. The amount of necessary
fine-tuning depends on the parameters of the superpotential,
especially on the ratio of the racetrack exponents $c$ and $d$.
In table \ref{tab_KL_polonyi} we present the dependence of
$\Delta_{\rm }$ (and some other quantities characterizing our model)
on $c$ keeping $d$ fixed. The ratio $|c/d|$ controls also the height
of the barrier which separates the Minkowski vacuum from the runaway
region of large values of the volume.
For values of $|c/d|$ significantly different from unity, the barrier
is too low to keep the inflaton in the region near the Minkowski vacuum.
Similar observation in the context of the model with non-SUSY uplifting
was done in \cite{bo2}. However, in contrast to the model with non-SUSY
uplifting, in the model with the Polonyi field, $\Delta_{\rm }$ does not
monotonically increase with the ratio $|c/d|$.
This can be understood as follows. The inflationary inflection point
at a positive value of $\phi$ does not exist for the ratio $|c/d|$ above some
critical value. On the other hand, the inflationary inflection point may occur
at negative value of $\phi$ for any value of the ratio $|c/d|$ (a set of
sections of the inflationary part of the potential in such a case is presented
in figure \ref{KL_polonyi_negphi}). However, inflation taking place in the
vicinity of the inflection point at $\phi<0$ ends in the Minkowski vacuum only
for some finite range of the ratio $|c/d|$. If the value of $|c/d|$ is too big
the barrier separating the Minkowski vacuum from the
region of infinite volume is too low to stop the inflaton. While for too small
values of $|c/d|$ there exists a barrier which separates the inflationary
inflection point from the Minkowski vacuum. Even though the parameter $W_0$ is
adjusted to arrange for a very flat inflection point at $\phi<0$, the slow-roll conditions are also satisfied for a relatively large range of positive values of $\phi$. In consequence, the inflaton evolves longer in the slow-roll phase as compared to the case of inflation starting at $\phi>0$. Since the slow-roll trajectory is longer, the inflection point can be less flat in order to obtain at least $60$ e-folds of inflation. This results in smaller fine-tuning of $W_0$ (i.e. bigger value of $\Delta$).
One can see this in table \ref{tab_KL_polonyi_negphi}. The value of $\Delta$ decreases when the ratio $|c/d|$ grows because the region at $\phi>0$ becomes less flat and more e-folds have to be generated very close to the inflection point at $\phi<0$.
Nevertheless, in any case inflation starting at $\phi<0$ requires less fine-tuning and $\Delta_{\rm }$ can be
even of the order $10^{-3}$. However, bigger values of
$\Delta_{\rm }$ are associated with bigger values of $\widehat{G}^2$.
This in turn implies that the necessary condition (\ref{etacon})
becomes more stringent which results in a smaller mass of $\theta$
field during inflation. For this mass being smaller than the Hubble
scale isocurvature perturbations would be produced.
This is not necessarily a problem because it depends on the details
of reheating whether this type of perturbations survive or not.
On the other hand, isocurvature perturbations can be a source for
adiabatic perturbations and can alter the value of the spectral
index in a way which is hard to guess without a detailed analysis
\cite{isocurvature}. This kind of analysis is beyond the scope of
this paper but it would be interesting to see whether the spectral
index would be still compatible with the WMAP data after taking into
account the isocurvature perturbations.

\subsubsection{Saddle point inflation with Polonyi uplifting}

We now investigate the saddle point model \cite{racetrack} with
the non-SUSY uplifting term (\ref{uplift}) replaced with the
Polonyi sector. For a numerical example we choose the following
set of parameters:
\begin{equation}
\begin{tabular}{lll}
$W_0=-\frac{1}{3260} , \qquad$ &
$C=\frac{8}{15} \,, \qquad$ &
$D=-\frac{14}{15} \,,$
\\[4pt]
$\mu^2=5.936801\cdot10^{-4} \, , \qquad$ &
$c=\frac{2\pi}{100} \, , \qquad$ &
$d=\frac{2\pi}{90} \, ,$
\end{tabular}
\label{paracetrack_polonyi}
\end{equation}
where $\mu^2$ is again adjusted to cancel the cosmological constant
in the post-inflationary vacuum. For the above set of parameters
inflation can take place in the vicinity of the saddle point located
at:
\begin{equation}
\tau_{\rm saddle}=0, \qquad \theta_{\rm saddle}=0,
\qquad t_{\rm saddle}\approx124.7, \qquad \phi_{\rm saddle}\approx0.2135.
\label{spP_inf}
\end{equation}
The $\eta$-matrix at this saddle point has three positive eigenvalues,
$\eta_2\approx4.49$, $\eta_3\approx40.3$, $\eta_4\approx45.9$,
and one negative, very small eigenvalue $\eta_1\approx-0.014$.
The unstable direction is
($t$, $\phi$, $\tau$, $\theta$)=(0, 0, -0.4565, 0.8897), so the inflaton
is a mixture of the imaginary parts of the fields with a somewhat bigger
contribution from the matter field. This is different from
the situation in the original model \cite{racetrack} in which
$\tau$ was the inflaton. SUSY breaking during
inflation is dominated by the matter field. The contribution from the
volume modulus is small, $\Theta_{T}^2|_{\rm inf}\approx0.16$, but about
one order of magnitude bigger than in the inflection point model of
subsection \ref{section_InfPol}. This can be understood as follows.
The inflationary inflection point arises from a minimum which was
supersymmetric before uplifting. The position of the inflection
point is very similar to the position of that minimum.
The volume modulus $F$-term vanishes at the SUSY minimum, so it is
also very small at the inflationary inflection point.
The situation is different in the case of a saddle point in which
SUSY is broken already before uplifting. Since the uplifting only
mildly affects the position of the saddle point, the volume modulus
$F$-term is comparable before and after uplifting. In consequence
the contribution of the volume modulus to SUSY breaking at the
inflationary saddle point is non-negligible.
A bigger contribution of the modulus field to the goldstino
direction results in a bigger value of the corresponding
sectional curvature. For the saddle point we obtained
$R(f^i)\approx0.018$ which is two orders of magnitude bigger that in
the inflection point model. Nevertheless, inflation in the present
model can be realized because $\widehat{G}^2$ has a relatively
small value during inflation. At the saddle point it is
$\widehat{G}^2|_{\rm inf}\approx3.8$, so $R(f^i)$ is significantly
smaller than $2/\widehat{G}^2$ during inflation and the condition
(\ref{etacon}) is satisfied.

The potential has two Minkowski vacua at:
\begin{equation}
\tau_{\rm vac}\approx\mp18.7, \qquad \theta_{\rm vac}\approx\pm0.0894,
\qquad t_{\rm vac}\approx102.5, \qquad \phi_{\rm vac}\approx0.643.
\label{spP_vac}
\end{equation}
Supersymmetry breaking at these minima is again dominated by the
matter field. This dominance in the vacuum is much stronger than during
inflation. The contribution from the volume modulus
at the vacuum, $\Theta_{T}^2|_{\rm vac}\approx0.017$, is an order
of magnitude smaller than at the saddle point,
$\Theta_{T}^2|_{\rm inf}\approx0.16$.
The reason is that the minima are supersymmetric before uplifting.
After uplifting, the positions of the minima are only mildly shifted
so the volume modulus $F$-term at the vacuum remains very small.

The mass spectrum at the Minkowski vacuum is quite similar to that
of the inflection point model of subsection \ref{section_InfPol}.
The matter-modulus mass matrix at the vacuum has the following
eigenvalues:
$m_{\phi}\approx10^{12} {\rm GeV}$,
$m_{\theta}\approx5\cdot10^{11} {\rm GeV}$,
$m_t\approx7\cdot10^{12} {\rm GeV}$,
$m_{\tau}\approx8\cdot10^{12} {\rm GeV}$.
The components of the matter field are about one order of
magnitude lighter than the volume modulus and the axion.
The gravitino mass at the Minkowski minima,
$m_{3/2}\approx6\cdot10^{11} {\rm GeV}$, is bigger than
the Hubble scale during inflation $H\approx10^{11} {\rm GeV}$.
The mass of the real component of the matter field is smaller
than the masses of the volume modulus and the axion
also at the inflationary saddle point:
$m^2_{\phi_{\rm inf}}\approx (5\cdot10^{11} {\rm GeV})^2$,
$m^2_{t_{\rm inf}}\approx m^2_{\tau_{\rm inf}}\approx(10^{12} {\rm GeV})^2$.
Of course, the inflaton (dominated by the imaginary component
of the matter field) is tachyonic:
$m^2_{\theta_{\rm inf}}\approx-(3\cdot10^{10} {\rm GeV})^2$.
All non-tachyonic scalar fields have during inflation masses larger
than the Hubble scale, so the isocurvature perturbations are strongly
suppressed in this model.

\begin{figure}[t]
\centering
  \includegraphics[width=8cm,height=6cm,angle=0]{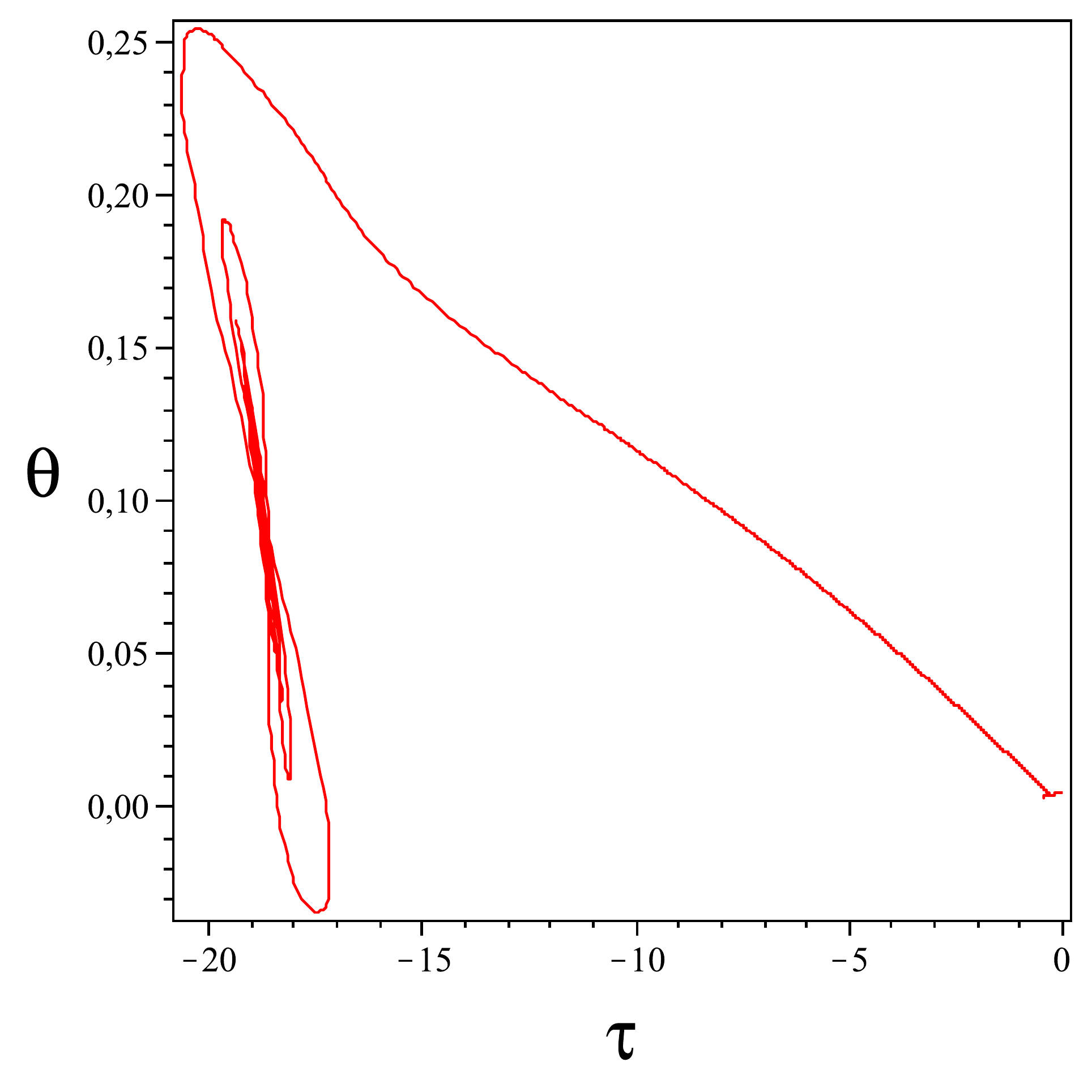}
  \includegraphics[width=8cm,height=6cm,angle=0]{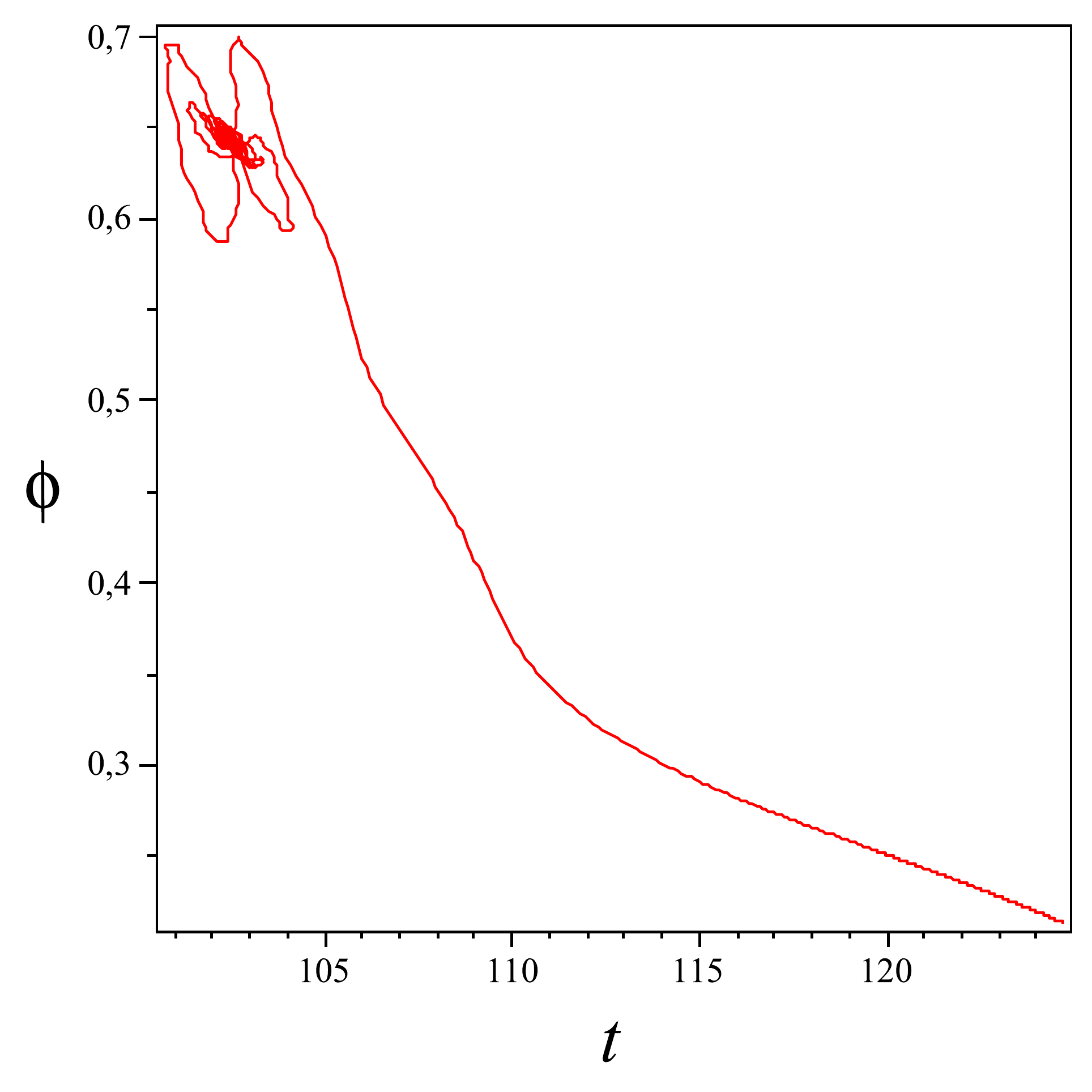}
  \caption{
Fields trajectory from the inflationary saddle point
to the Minkowski minimum in the model with the parameters
(\ref{paracetrack_polonyi}).}
\label{race_traj}
\end{figure}

Comparing (\ref{spP_inf}) with (\ref{spP_vac}) one can see that
the values of all the fields are different during and after inflation.
This is the reason why it is not possible to plot the global picture
of the potential. Therefore, in figure \ref{race_traj} we show the
inflationary trajectories of all the fields. These trajectories were
obtained by solving numerically equations of motion (\ref{eqmot})
with the fields set initially at rest close to the saddle point:
$\theta_{\rm ini}=0.005$, $\tau_{\rm ini}=0$,
$t_{\rm ini}=t_{\rm saddle}$, $\phi_{\rm ini}=\phi_{\rm saddle}$.
With these initial conditions inflation lasts about $77$ e-folds and
ends in one of the Minkowski minima.

The parameters (\ref{paracetrack_polonyi}) satisfy the COBE
normalization and give the amplitude of the density perturbations
$\frac{\delta\rho}{\rho}\approx2\cdot 10^{-5}$. The spectral index
about $55$ e-folds before the end of inflation has a value
$n_s\approx0.95$, which is very similar to that found in the original
model \cite{racetrack} with non-SUSY uplifting. We computed the
value of the spectral index for many various choices of the parameters
and found that it is in a good agreement (at the level of one
per mill) with the following approximate analytic formula developed
in \cite{lw,nsbrax}:
\begin{equation}
\label{ns_saddle}
n_s\approx1 + 2\eta_{\rm sad}-\frac{6\eta_{\rm sad}}{1-e^{-2N_e\eta_{\rm sad}}}
\,,
\end{equation}
where $\eta_{\rm sad}$ is the value of the slow-roll parameter
$\eta$ at the saddle point while $N_e$ is the number of e-folds
between the time when a given scale crosses the horizon and the
end of inflation. The above formula was derived under the assumption
that inflation is dominated by one field, say $x$, and the potential
is invariant under the transformation $x\rightarrow -x$. In the present
model the inflaton is a combination of two fields, $\theta$ and $\tau$,
but the potential is invariant under the transformation:
$\theta\rightarrow -\theta$, $\tau\rightarrow -\tau$. This is the reason
why the formula (\ref{ns_saddle}) works well. Therefore, the bound
$n_s\lesssim0.95$ which follows from (\ref{ns_saddle}) with $N_e=55$
is indeed the robust prediction of the saddle point racetrack inflationary
models independently of the form of uplifting. We also found unmeasurably
small values of the tensor to scalar perturbation amplitudes ratio
$r\approx10^{-7}$ and the running of the spectral index
$\frac{d n_s}{d\ln k}\approx10^{-3}$.

In order to obtain the spectral index compatible at the $3\sigma$ level
with the 5-year WMAP data \cite{wmap} one needs fine-tuning of the
parameter $W_0$ at the level $\Delta_{\rm }\approx2\cdot10^{-3}$.
Thus, the part of the parameter space allowing for phenomenologically
acceptable inflation
is bigger than in the original model \cite{racetrack} with
non-SUSY uplifting in which $W_0$ has to be fine-tuned at the level of
$10^{-4}$. We checked also that the required fine-tuning is
comparable for different sets of parameters of the superpotential.

We end this subsection with the conclusion that the Polonyi field can
be considered as a source of uplifting and spontaneous SUSY breaking
in inflationary models based on the racetrack superpotential.
In such models the Polonyi field is usually lighter than the volume
modulus and its imaginary part becomes the main component of
the inflaton. The main advantage of
these models as compared to models with non-SUSY uplifting is that
the fine-tuning required for successful inflation is reduced,
especially in the case of the inflection point inflation.

\subsection{O'uplifting}
\label{ouplsec}

Now we turn our attention to models with the following K\"ahler
potential for the matter field:
\begin{equation}
\label{okahler}
K^{(\Phi)}=\Phi\ov{\Phi}-\frac{(\Phi\ov{\Phi})^2}{\Lambda^2}
\,.
\end{equation}
A model described by the above K\"ahler potential and the superpotential
(\ref{Wmatter}) with $c_0=0$ can be treated as an effective quantum
corrected O'Raifeartaigh model \cite{O'Raifeartaigh,okklt} with the superpotential
\begin{equation}
\label{WO}
W^{(O')}=mXY+\lambda\Phi X^2-\mu^2\Phi
\,,
\end{equation}
 in which the fields $X$ and $Y$
are assumed to be heavy and are integrated out \cite{okklt}.
The correction to the canonical K\"ahler potential originates from
the expansion of the one loop expression
$K^{\rm{1}-loop}_{O'}=\Phi\ov{\Phi}
\left(1-\frac{c_1\lambda^2}{16\pi^2}\log\left(1
+\frac{\lambda^2\Phi\ov{\Phi}}{m^2}\right)\right)$ for
$\frac{\lambda^2\Phi\ov{\Phi}}{m^2}\ll 1$, where parameter $c_1$ is of
order one. Comparing this expansion with (\ref{okahler}) one obtains
$\Lambda^2=\frac{16\pi^2m^2}{c_1\lambda^4}$.
In order to guarantee that the expression for the one loop K\"ahler
potential is a good approximation, one has to assume that
$m^2\gg\lambda\mu^2$ \cite{okklt}.
This condition also assures that the fields $X$ and $Y$ vanish at
the minimum of the resulting potential. In addition it is assumed
that $\frac{\lambda}{16\pi^2}\ll1$ and $\Lambda^2\ll1$.
The authors of \cite{okklt} pointed out that one can use such an
effective O'Raifeartaigh model to uplift the KKLT model and called that
scenario O'KKLT. They showed that all of the above consistency
conditions can be fulfilled in such model.

The K\"ahler metric for the K\"ahler potential (\ref{okahler}) is given
by $K_{\Phi\ov{\Phi}}=1-4\Phi\ov{\Phi}/\Lambda^2$. It is positive definite
only for $|\Phi|<\Lambda/2$. Therefore, for $\Lambda\ll1$ a model
with the K\"ahler potential (\ref{okahler}) can be well defined
only for $|\Phi|\ll1$. The curvature scalar of the matter field manifold
equals $R_{\Phi}=-4/\left[\Lambda^2(1-4|\Phi|^2/\Lambda^2)^3\right]$.
Notice that this curvature is negative in the region where the K\"ahler
metric is positive. So, in the case in which the matter field
dominates SUSY breaking the necessary condition (\ref{etacon}) can be
satisfied independently of the value of $\widehat{G}^2$.
Thus, by coupling this model to the volume modulus one can improve
stability of the Minkowski vacuum and increase the upper bound
(\ref{etamax}) on the parameter $\eta$ in a more efficient way
(i.e.\ with a smaller contribution of the matter field to SUSY breaking)
than in the case of a canonically normalized matter field.

In \cite{Kitano} a model with the K\"ahler potential (\ref{okahler})
and the superpotential (\ref{Wmatter}) was proposed.
It was shown that for an appropriate choice of the parameters a SUSY
breaking Minkowski minimum can exist. In the limit $\Lambda\ll1$ such a
minimum occurs at
\begin{equation}
\phi_{\rm min}\approx\frac{\sqrt{3}}{6}\Lambda^2 ,
\qquad\qquad \theta_{\rm min}=0 \,,
\end{equation}
i.e\ at a very small value of the matter field: $\phi_{\rm min}\ll\Lambda\ll1$.
It was found in \cite{okklt} that the position of this Minkowski minimum is
not significantly shifted after coupling that model to the volume modulus.
Notice that after integrating out the heavy fields $X$ and $Y$ from the full
O'Raifeartaigh model with the superpotential $W^{(O')}$ (\ref{WO}), the
effective superpotential is given by $W^{(\Phi)}$ (\ref{Wmatter}) with the
vanishing constant term $c_0$ (however, the constant term originating from
fluxes is still present in $W^{(T)}$).

In the limit $\Lambda\rightarrow\infty$, K\"ahler potential (\ref{okahler})
reduces to the canonical one used in the Polonyi model discussed before.
However, we are interested now in the case $\Lambda\ll1$.
In contrast to the Polonyi model, the value of the
field $\Phi$ should be much below the Planck scale $\phi\ll\Lambda\ll1$.
This hierarchy is a source of some features of the mass matrix
of the modulus-matter system which will prove to be very important
for the inflationary dynamics.
Before discussing these features it should be stressed that they arise not only for the racetrack superpotential (\ref{Wrace}) but for any generic superpotential $W^{(T)}$ describing the modulus sector.
The diagonal entries of the mass matrix in the matter field sector are
enhanced because $m^2_{\phi\phi}, m^2_{\theta\theta}\sim\Lambda^{-2}$.
The diagonal entries in the volume modulus sector, $m^2_{tt}$ and
$m^2_{\tau\tau}$, as well as the off-diagonal
entry $m^2_{\tau\theta}$ do not depend on $\Lambda$ in the leading order.
The off-diagonal entry $m^2_{t\phi}$ does depend on $\Lambda$ but one can
show that the ratio $m^2_{t\phi}/m^2_{\phi\phi}\approx-2\sqrt{3}\phi\ll1$.
All other off-diagonal entries vanish for $\tau=\theta=0$ (this is
true during inflation also for all other inflationary racetrack models
considered in this paper). Thus, the mass matrix is nearly diagonal
and there is negligible mixing between the volume modulus and the matter
field. In addition, the matter field is heavier than the volume modulus
so it is possible to approximately decouple the matter field from the
inflationary dynamics. Nonetheless, one cannot neglect the matter
field completely because it contributes to SUSY breaking. This is very
desirable because the matter field $F$-term provides an effective uplifting
term $e^{K^{(\Phi)}}|D_{\Phi}W|^2/(T+\ov{T})^3$ in the scalar potential.
The magnitude of the matter field $F$-term is controlled by the parameter $\mu^2$ so one can use this parameter to cancel the cosmological constant in the post-inflationary vacuum.
In addition, the contribution of the matter field to SUSY breaking
is important for the stability of the model because it reduces the
sectional curvature along the direction of SUSY breaking $R(f^i)$.
As we shall see in the following examples, after coupling the racetrack
model to an effective O'Raifeartaigh model, $R(f^i)$ can be small enough
to satisfy the constraint (\ref{etacon}) at the Minkowski minimum
as well as in the inflationary region.

\subsubsection{Inflection point inflation with O'uplifting}

\begin{figure}
  \centering
  \includegraphics[width=8cm,angle=0]{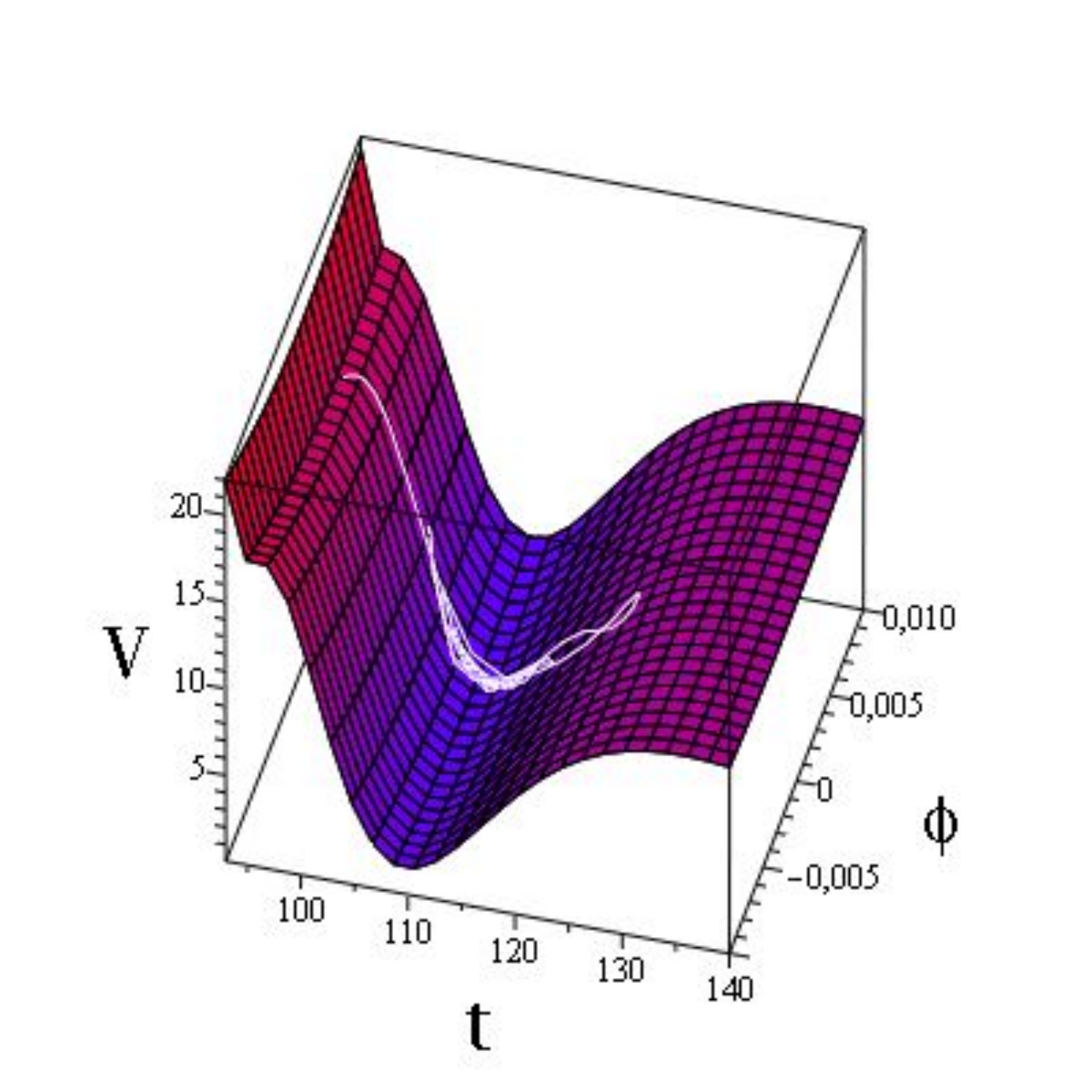}
  \caption{
The potential multiplied by $10^{21}$ for the inflection point model
coupled to an effective O'Raifeartaigh model with the parameters
(\ref{parKL_okklt}) and $\tau=\theta=0$. The white curve represents
the field trajectory from the inflationary inflection point to the
Minkowski vacuum.}
  \label{KL_okklt}
\end{figure}

We first apply the just discussed O'uplifting to a model in which
inflation takes place in the vicinity of an inflection point.
The superpotential is given by the sum of (\ref{Wrace}) and
(\ref{Wmatter}) while the K\"ahler potential is the sum of
(\ref{kahler}) and (\ref{okahler}).
For a numerical example we choose the following set of parameters:
\begin{equation}
\begin{tabular}{llll}
$W_0=-3.426786982\cdot10^{-7} , \qquad$ &
$C=-\frac{2}{5} \,, \qquad$ &
$D=\frac{1}{200} \,,$ &
\\[4pt]
$\mu^2=5.168125\cdot10^{-7} \, , \qquad$ &
$c=\frac{2\pi}{40} \, , \qquad$ &
$d=\frac{2\pi}{60} \, , \qquad$ & $\Lambda=\frac{1}{10}$ \,,
\end{tabular}
\label{parKL_okklt}
\end{equation}
where again $W_0=A+c_0$. The resulting potential has an inflection
point and a Minkowski minimum both at $\tau=\theta=0$, so the
imaginary parts of the fields can vanish during and after inflation.
Thus, the inflationary dynamics becomes effectively 2-dimensional
(if we do not use initial conditions with non-zero values or
velocities in $\tau$ and $\theta$ directions). The potential in
the ($t$-$\phi$) plane is shown in figure \ref{KL_okklt}.
The inflationary inflection point is situated at $t_{\rm inf}\approx95.69$,
$\phi_{\rm inf}\approx3.097\cdot10^{-3}$. The field $t$ plays the role
of the inflaton as in the original model \cite{lw} with non-SUSY
uplifting. A new feature is that in the present model SUSY breaking
during inflation is dominated by the matter field:
$\Theta_{\Phi}^2|_{\rm inf}\approx0.967$.
The curvature in the direction of SUSY breaking
is large and negative, $R(f^i)\approx-372$, so the
condition (\ref{etacon}) necessary for inflation is satisfied
independently of the value of $\widehat{G}^2$.
All 3 non-zero eigenvalues of the $\eta$-matrix are positive and very
large: $\eta_2\approx541$, $\eta_3\approx1045$, $\eta_4\approx1061$.
The second slow-roll parameter is extremely small:
$\epsilon\approx2 \cdot10^{-17}$. This implies that the scale of
inflation is a few orders of magnitude smaller than in the previously
discussed models with a canonically normalized matter field.
Indeed, we obtain $H\approx2\cdot10^8 {\rm GeV}$.
Such scale of inflation is still much too high to admit low
energy SUSY breaking. As expected, the gravitino mass is
of the same order as $H$: $m_{3/2}\approx2\cdot10^8 {\rm GeV}$.

In order to get more predictions of the model we solved numerically the
equations of motion (\ref{eqmot}) and obtained the following results.
Inflation lasts about $176$ e-folds. As seen from figure \ref{KL_okklt},
the barrier is high
enough to stop the inflaton which eventually settles in the Minkowski
vacuum at $t_{\rm vac}\approx109.2$, $\phi_{\rm vac}\approx2.12\cdot10^{-3}$.
Matter field contribution to SUSY breaking at this minimum is very
similar to the corresponding contribution during inflation
$\Theta_{\Phi}^2|_{\rm vac}\approx\Theta_{\Phi}^2|_{\rm inf}\approx0.967$.
The spectral index at the COBE normalization scale is
$n_s\approx0.93$ and is well approximated by the formula (\ref{fns})
for other sets of parameters. The values of the tensor
to scalar perturbation ratio and of the running of
the spectral index are very small: $r\approx10^{-12}$,
$\frac{d n_s}{d\ln k}\approx10^{-3}$.

We argued earlier that in this model the volume modulus should be
lighter than the matter field during and after inflation. We found
the following mass eigenvalues during slow-roll inflation:
$m^2_{t_{\rm inf}}\approx -(6\cdot10^{7} {\rm GeV})^2$,
$m^2_{\tau_{\rm inf}}\approx(7\cdot10^{9} {\rm GeV})^2$,
$m^2_{\phi_{\rm inf}}\approx(10^{10} {\rm GeV})^2$,
$m^2_{\theta_{\rm inf}}\approx(10^{10} {\rm GeV})^2$,
and at the Minkowski minimum:
$m^2_t\approx m^2_{\tau}\approx(3\cdot10^{9} {\rm GeV})^2$,
$m^2_{\phi}\approx m^2_{\theta}\approx(7\cdot10^{9} {\rm GeV})^2$.
As expected, the volume modulus is lighter but the difference
is not big.

Our model does not significantly differ from the original one
proposed in \cite{lw} except the fact that SUSY in the
post-inflationary vacuum is broken spontaneously by the
matter field. The unstable direction at the inflection point is
given by the vector
($t$, $\phi$, $\tau$, $\theta$)=(0.999, -0.051, 0, 0)
and has only marginal contribution from the matter field.
So, $t$ is the inflaton as in the original model.
This decoupling of the matter field follows from the strong suppression
of the off-diagonal entries in the mass matrix. The only non-vanishing
off-diagonal entries are: $m^2_{\tau\theta}\approx m^2_{t\phi}
\approx -(2\cdot10^{9} {\rm GeV})^2$ during the slow-roll phase
of inflation and
$m^2_{\tau\theta}\approx m^2_{t\phi}\approx (10^{9} {\rm GeV})^2$
at the Minkowski vacuum.
These off-diagonal entries are significantly smaller than the trace
of the mass matrix. The suppression is even stronger for smaller
values of $\Lambda$.

One of the consequences of the decoupling of the matter field is
that the solution of the overshooting problem requires extreme
fine-tuning as in the original model \cite{lw} with non-SUSY uplifting.
The parameters ($\ref{parKL_okklt}$) were chosen in such a way that
the height of the barrier is just enough to prevent the inflaton
from running away to infinite volume. The fine-tuning of the parameter
$W_0$ required for successful inflation is found to be
$\Delta_{\rm }\approx10^{-9}$. Thus, the fine-tuning is even stronger
than in the original model \cite{lw} with non-SUSY uplifting from
$\ov{D3}$-branes. The difference follows from the fact that uplifting
from $\ov{D3}$-branes is proportional to $1/t^2$ while effective
uplifting from the matter field scales like $1/t^3$. Therefore,
effective uplifting from matter field decreases faster with $t$ and
in consequence the height of the barrier is smaller. The barrier
can be made higher only when $C/D$ is made larger and/or $|c-d|$
is made smaller. However, this kind of change of parameters results
in stronger fine-tuning required for phenomenologically
acceptable inflation, as was explained in \cite{bo2}.

\subsubsection{Saddle point inflation with O'uplifting}

\begin{figure}
  \centering
  \includegraphics[width=8cm,angle=0]{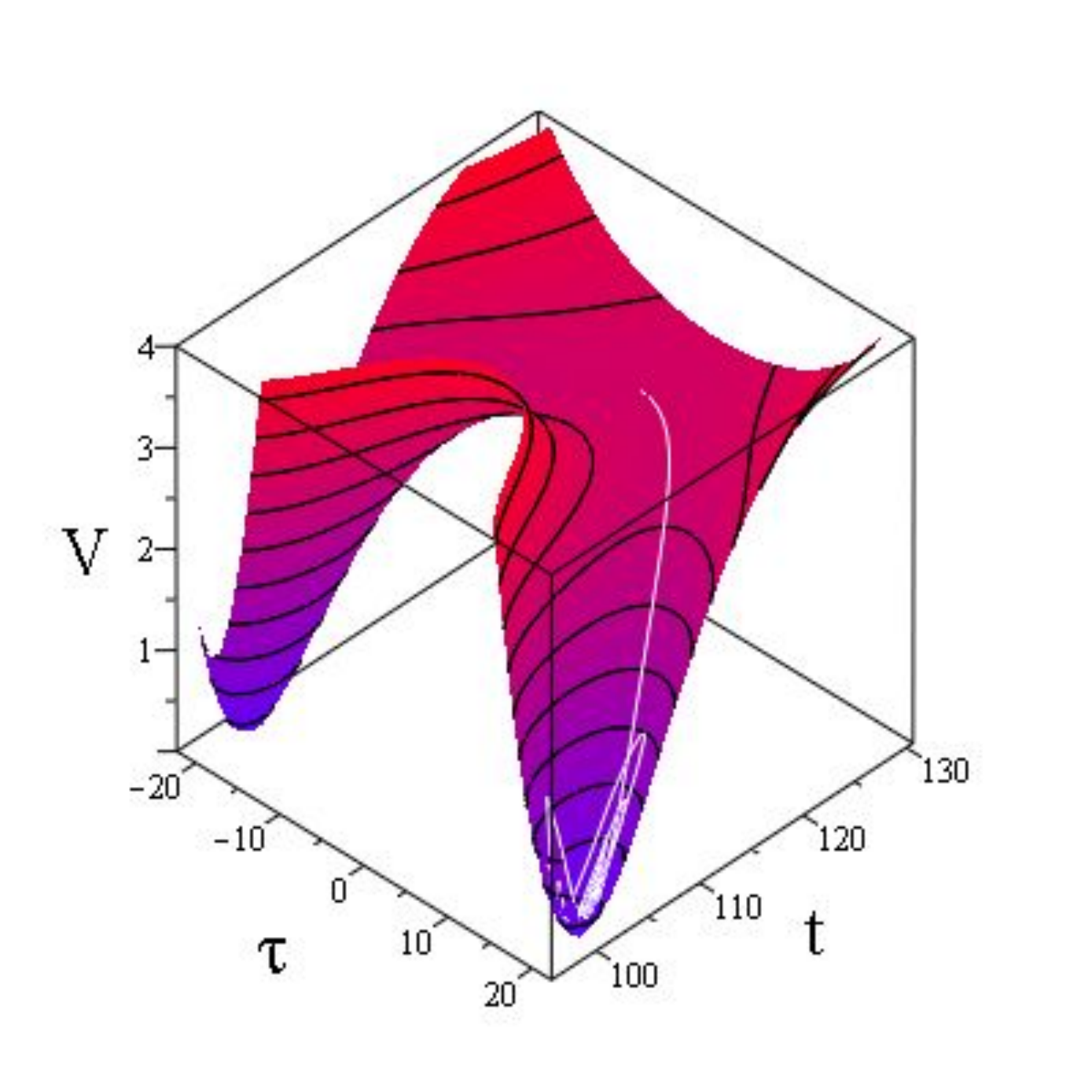}
  \caption{
The potential multiplied by $10^{16}$ for the saddle point model
coupled to an effective O'Raifeartaigh model with the parameters
(\ref{paracetrack_okklt}). The matter field is fixed at $\phi=\phi_{\rm vac}$,
$\theta=\theta_{\rm vac}$. The white curve represents the field trajectory from
the inflationary saddle point to the Minkowski vacuum.}
  \label{racetrack_okklt}
\end{figure}

Let us use an effective O'Raifeartaigh model for uplifting in the
racetrack model with an inflationary saddle point.
The superpotential and the K\"ahler potential have the
same form as in the previous subsection but the values
of the parameters are different.
For the numerical example we choose the following set:
\begin{equation}
\begin{tabular}{llll}
$W_0=-\frac{1}{13005} , \qquad$ &
$C=\frac{1}{20} \,, \qquad$ &
$D=-\frac{7}{80} \,,$ &
\\[4pt]
$\mu^2=1.073049\cdot10^{-4} \, , \qquad$ &
$c=\frac{2\pi}{100} \, , \qquad$ &
$d=\frac{2\pi}{90} \, , \qquad$ & $\Lambda=\frac{1}{10}$ \,.
\end{tabular}
\label{paracetrack_okklt}
\end{equation}
The potential as a function of $t$ and $\tau$
for small fixed values of $\phi$ and $\theta$
is shown in figure
\ref{racetrack_okklt} and looks similar to the potential in the
original model \cite{racetrack} with non-SUSY uplifting (shown in
figure \ref{racetrack3d}). The inflationary saddle point is located at:
\begin{equation}
\tau_{\rm saddle}=0, \qquad \theta_{\rm saddle}=0,
\qquad t_{\rm saddle}\approx122.4,
\qquad \phi_{\rm saddle}\approx-2.394\cdot 10^{-4}.
\end{equation}
The $\eta$-matrix at this saddle point has 3 positive
eigenvalues: $\eta_2\approx18.3$, $\eta_3\approx1126$,
$\eta_4\approx1126$, and one very small, negative one:
$\eta_1\approx-0.000282$. The inflaton direction
($t$, $\phi$, $\tau$, $\theta$)=(0, 0, 0.9999, -0.0066) is very
strongly dominated by the axion $\tau$ as in the original racetrack
model \cite{racetrack} with non-SUSY uplifting.
In contrast to the models considered in the previous subsections,
the contributions of the matter field and the volume modulus to SUSY
breaking at the saddle point are comparable:
$\Theta_{\Phi}^2|_{\rm inf}\approx0.6$, $\Theta_{T}^2|_{\rm inf}\approx0.4$.
Nevertheless, the sectional curvature along the SUSY breaking direction,
$R(f^i)\approx-143$, is negative and the necessary condition for
slow-roll inflation (\ref{etacon}) is satisfied independently of
the value of $\widehat{G}^2$.

There are two Minkowski minima situated at:
\begin{equation}
\tau_{\rm vac}\approx\mp19.7, \qquad \theta_{\rm vac}\approx\pm1.78\cdot10^{-4},
\qquad t_{\rm vac}\approx100.4, \qquad \phi_{\rm vac}\approx2.25\cdot10^{-3}.
\end{equation}
SUSY breaking at these minima is dominated by the matter field:
$\Theta_{\Phi}^2|_{\rm vac}\approx0.962$.
The eigenvalues of the mass matrix are as follows:
$m^2_t\approx(7\cdot10^{11} {\rm GeV})^2$,
$ m^2_{\tau}\approx(8\cdot10^{11} {\rm GeV})^2$,
$m^2_{\phi}\approx m^2_{\theta}\approx(2\cdot10^{12} {\rm GeV})^2$.
The volume modulus is a bit lighter than the matter field in contrast
to the saddle point model coupled to the Polonyi field. The gravitino
mass at the Minkowski minima is $m_{3/2}\approx5\cdot10^{10} {\rm GeV}$
and, as expected, exceeds the Hubble scale during
inflation $H\approx2\cdot10^{10} {\rm GeV}$.
The mass spectrum at the inflationary saddle point is similar to that at the vacuum:
$m^2_{\tau}\approx-(7\cdot10^{8} {\rm GeV})^2$,
$m^2_t\approx(2\cdot10^{11} {\rm GeV})^2$,
$m^2_{\phi}\approx m^2_{\theta}\approx(10^{12} {\rm GeV})^2$. The mass of the second lightest scalar field during inflation is much larger than the Hubble scale so this model does not predict generation of substantial isocurvature perturbations.

We solved numerically equations of motion (\ref{eqmot}) with the fields
initially set at rest close to the saddle point:
$\tau_{\rm ini}=0.1$, $\theta_{\rm ini}=0$,
$t_{\rm ini}=t_{\rm saddle}$, $\phi_{\rm ini}=\phi_{\rm saddle}$.
The trajectory for the volume modulus is shown in figure
\ref{racetrack_okklt}. Inflation lasts about $303$ e-folds. After
inflation fields roll down towards one of the Minkowski minima where
they eventually settle down after a period of oscillations.

We checked that the spectral index is well approximated by the analytic
formula (\ref{ns_saddle}). For the parameters (\ref{paracetrack_okklt})
its value $55$ e-folds before the end of inflation is
$n_s\approx0.94$. As in the previous models discussed
in this paper, the tensor to scalar perturbations ratio and the running
of the spectral index are very small. In this case $r\approx10^{-9}$,
$\frac{d n_s}{d\ln k}\approx10^{-3}$.

The success of inflation relies on an appropriate choice of the
parameter $W_0$. For the parameters (\ref{paracetrack_okklt}) necessary
fine-tuning is $\Delta_{\rm }\approx6\cdot10^{-4}$ so it is comparable
to the original model \cite{racetrack} with non-SUSY uplifting.
We checked that the value of $\Delta_{\rm }$ is not sensitive
to changes of other parameters.

\subsubsection{Consistency conditions}

In order to treat a model with the K\"ahler potential (\ref{okahler})
as an effective quantum corrected O'Raifeartaigh model, some consistency
conditions have to be satisfied. They can be summarized as follows:
\begin{equation}
|\Phi|^2\ll\frac{m^2}{\lambda^2}, \qquad m^2\gg\lambda\mu^2,
\qquad m,\mu,\frac{\lambda}{4\pi}\ll1.
\label{conscond}
\end{equation}
In our examples the value of $|\Phi|^2$ during and after inflation does not exceed $10^{-5}$.
Hence, assuming for a moment that $\lambda$ is of order one, the first
condition in (\ref{conscond}) implies $m^2\gg10^{-5}$. The second condition
is more restrictive because both models considered in this section
have a high scale of inflation and in consequence $\mu^2$ is rather large.
In the saddle point model $\mu^2\sim10^{-4}$ while in the inflection
point model, in which the scale of inflation is a bit lower,
$\mu^2\sim10^{-6}$. This implies that $m^2$ should be rather large
but still well below the Planck scale, as required by the third
condition in (\ref{conscond}). Once $m$ and $\Lambda$ are fixed,
$\lambda$ is no longer a free parameter because
$\lambda^2=\frac{4\pi m}{\Lambda\sqrt{c_1}}$.
This implies that also $\Lambda$ should have a relatively big value.
Since $m^2$ cannot be too small because of the second condition in
(\ref{conscond}), a very small value of $\Lambda$ would lead to the
violation of the condition $\lambda\ll4\pi$. As an example of
a consistent set of parameters we chose $m^2=10^{-3}$ and
$\Lambda=10^{-1}$. For $c_1=1$, one obtains
$\lambda\approx2$ which is indeed of order one and also still much
smaller than $4\pi$. Therefore, all consistency conditions can be
satisfied.

To summarize, both models under consideration can be uplifted by an
effective quantum corrected O'Raifeartaigh model provided that $\Lambda$
and $m^2$ are not too small (but for consistency they must be much
below 1). In these models the matter field is heavier than the volume
modulus and the mixing between the matter field and the volume
modulus is suppressed. Therefore, the matter field is almost decoupled
from the inflationary dynamics and most of the features present in the
original models \cite{lw} and \cite{racetrack} with non-SUSY uplifting
are recovered.

\section{Inflation and the scale of SUSY breaking}
\label{scaleSUSYsec}

In all models discussed in the previous section the gravitino mass
in a post-inflationary vacuum is much bigger than the electroweak
scale. This is a generic feature of string inspired inflationary
models. It was pointed out in \cite{kl} that in typical models based
on the KKLT moduli stabilization the value of the gravitino mass
exceeds the Hubble scale during inflation. Most of string motivated
models constructed so far, including those discussed in the previous
section, have the scale of inflation much bigger than the electroweak
scale. Therefore all of these models predict the energy scale of SUSY
breaking inaccessible at the LHC.

The possible solution to this problem was suggested in \cite{kl}.
If inflation ends in a SUSY Minkowski minimum the gravitino mass
vanishes independently of the scale of inflation. However, in that
case there is no uplifting (or it is very small). The tree-level
K\"ahler potential for the volume modulus (\ref{kahler}) violates
necessary condition (\ref{RTcon}), so inflation cannot be realized,
as was shown in \cite{bo}. It was proposed in \cite{bo} that the
condition (\ref{RTcon}) can be satisfied once $\alpha'$ \cite{alphaprime} and/or
string loop \cite{stringloop} corrections are added to the K\"ahler potential.
With these corrections successful models of inflation with a TeV
range gravitino mass were constructed. The first model proposed
in \cite{bo} contains three non-perturbative terms in the
superpotential. In \cite{bo2} it was found that the form of the
superpotential, allowing for a successful inflation, can be
simplified when one admits positive exponents in the gaugino condensation
terms. With that assumption two gaugino condensation terms or even
one such term with threshold corrections are enough to accommodate
high scale inflation and a low scale of SUSY breaking \cite{bo2}.

In the one-field case, investigated in \cite{bo,bo2}, adding the
string corrections to the K\"ahler potential results in an increased
trace of the $\eta$-matrix. For large enough corrections the trace
of the $\eta$-matrix can become positive, so inflation can be realized.
In other words, the string corrections to the K\"ahler potential
stabilize the direction orthogonal to the inflaton (there is just
one such direction in the one complex field case). Since
the string corrections affect only the K\"ahler potential, the SUSY
Minkowski minimum remains unaffected by these corrections. Therefore,
including the string corrections to the K\"ahler potential is a
generic mechanism stabilizing the direction orthogonal to the inflaton.

\subsection{Seeking volume modulus inflation}

We address now the question if inflation ending in a SUSY Minkowski
minimum can be realized when the volume modulus is coupled to the
matter field and no string corrections to the K\"ahler potential are
included. Let us first concentrate on the case in which the volume
modulus is the main component of the inflaton field. In principle,
one could hope that the matter field will play a similar role to the
one played by the string corrections in the one-field case. Namely,
it may modify sectional curvature in the Goldstino direction $R(f^i)$
at the inflationary region in such a way that the condition (\ref{etacon})
can be satisfied. However, in this case things are much more
complicated since the matter field enters not only the K\"ahler
potential but also the superpotential. Moreover, in order to modify
$R(f^i)$ the matter field has to contribute to SUSY breaking at the
inflationary region. For example, for a canonically normalized matter
field the condition necessary for inflation (\ref{thetacon})
gives the following lower bound on the matter field contribution
to SUSY breaking:
\begin{equation}
\Theta_{\Phi}^2 > 1 - \frac{\sqrt{3}}{\widehat{G}}
\,.
\end{equation}
Typical values of $\widehat{G}$ are significantly larger than $\sqrt{3}$.
In all models proposed in \cite{bo} and \cite{bo2} the value of
$\widehat{G}^2$ at the inflationary point is of order $10$.
Therefore, the $F$-term  for the matter field at the inflationary
region should be comparable or even larger than the corresponding
$F$-term for the volume modulus. At the same time, contribution of
the matter field to the SUSY breaking at the minimum should be strongly
suppressed. This is because the starting point (before including the
matter field) is a minimum with (almost) vanishing cosmological constant.
Therefore, large  positive contribution of the matter field $F$-term to
the potential would generate a large positive cosmological constant at
the minimum. This is obviously not desirable from the phenomenological
point of view. In order to have so different values of the
$F$-terms at the inflationary point and at the minimum, one should
use a very specific form of the superpotential for the matter field.
In the previous section we investigated models with the linear
superpotential (\ref{Wmatter}) for the matter field. We have learned
from that analysis that for such superpotential the matter field
contributions to the SUSY breaking at the minimum and at the
inflationary point are comparable. This is what we want to avoid in
the present case. The linear superpotential for the matter field
is not appropriate, so we assume a quadratic one:
\begin{equation}
\label{Wmatter2}
W^{(\Phi)}=\mu^2(\Phi-\Phi_0)^2
\end{equation}
For the above superpotential, a solution to the SUSY preserving
condition $\pa W^{(\Phi)}/\pa \Phi\equiv W^{(\Phi)}_{\Phi}=0$
exists at $\Phi=\Phi_0$. In the following
we investigate a few models with the above superpotential
and different forms of the matter field K\"ahler potential.

First we consider the K\"ahler potential (\ref{okahler})
of the quantum corrected O'Raifeartaigh model with
$\Lambda\ll1$. The matter field $F$-term in this case is
$F_{\Phi}\sim[W^{(\Phi)}_{\Phi}+\ov{\Phi}(1-2|\Phi|^2/\Lambda^2)W]$.
In the limit $\Lambda\ll1$ the matter field $\Phi$ is heavier
than the volume modulus and remains approximately constant during
and after inflation.
Therefore, if $W^{(\Phi)}_{\Phi}$ vanishes at the minimum then it
is also very small at the inflationary region and does not contribute
to SUSY breaking. Moreover, the K\"ahler metric is positive definite
only for $|\Phi|<\Lambda/2\ll1$, so the second term in $F_{\Phi}$
is also very small. Hence, we conclude that in that case the matter
field does not significantly contribute to SUSY breaking at
the inflationary region and inflation cannot proceed without the
string corrections.

Let us now turn our attention to the case of the K\"ahler potential
(\ref{kahlercan}) which gives canonical kinetic terms for the matter
fields. In the previous section we have seen that in such a case,
in models with the vacuum obtained by uplifting a SUSY AdS minimum,
the inflaton is no longer the volume modulus alone but rather a
mixture of it with the matter field.
One may expect that also in models with inflation ending in a SUSY
Minkowski minimum, a canonically normalized matter field may play an
important role in the inflationary dynamics. To verify this expectation
we studied models of inflection point inflation proposed in \cite{bo2}
coupling them to a matter field instead of using the string corrections
to the K\"ahler potential.
It occurred that whether inflation ending in a SUSY Minkowski minimum
can be realized depends on details of a model. We constructed a model
based on the racetrack superpotential (\ref{Wrace}) with two positive
exponents\footnote{
We use different convention for the signs of the parameters $c$ and $d$.
In the notation of the present paper positive exponents correspond to
negative values $c$ and $d$.
}.
However, we were not able to construct any inflationary model
with one negative exponent in the racetrack superpotential
(\ref{Wrace}) in contrast to a corresponding model with the string
corrections and without the matter field \cite{bo2} (inflation in the model with two negative exponents cannot be realized even with the string corrections \cite{bo}). Moreover, in a model
in which inflation is possible (the one with two positive exponents)
there are some restrictions on its parameters.
Below we present this model in some detail.

\begin{figure}[t]
\centering
  \includegraphics[width=8cm,angle=0]{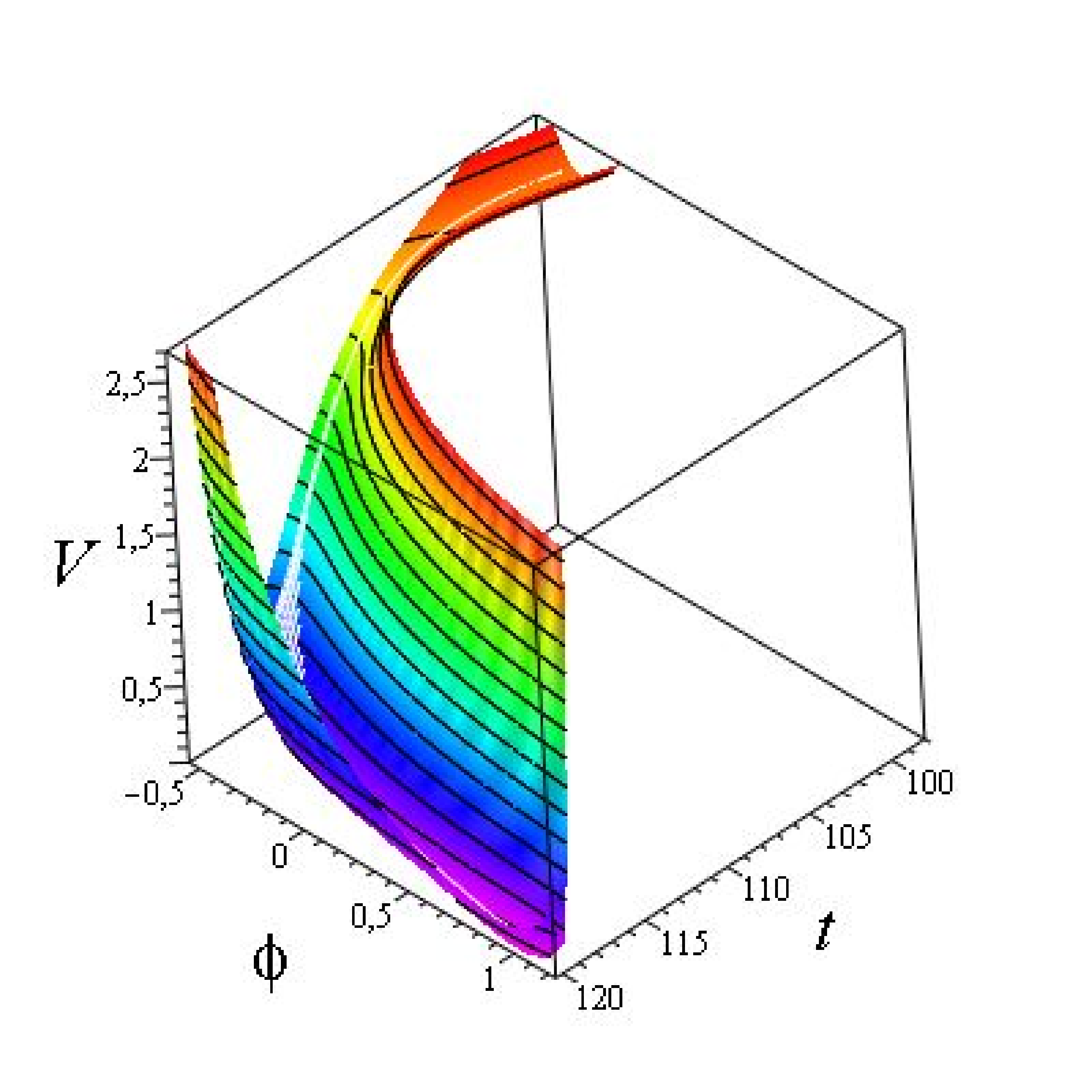}
  \caption{
The potential (\ref{2pospot}) multiplied by $10^{15}$ for the model
    defined by the parameter set (\ref{par2pos}). The white curve represents
    the inflationary trajectory from the inflection point to the SUSY (near)
    Minkowski minimum.}
\label{2pospotfig}
\end{figure}

\begin{figure}[t]
\centering
  \includegraphics[width=17cm,angle=0]{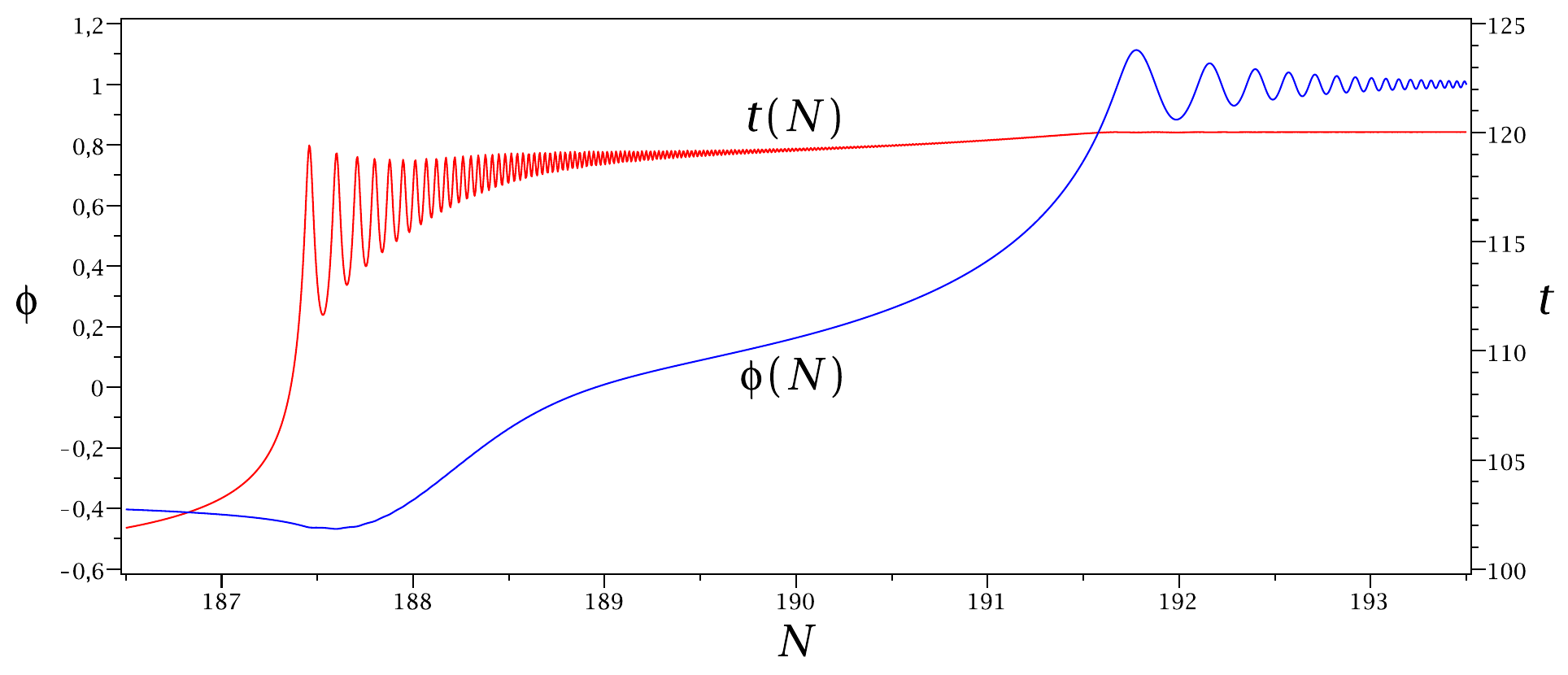}
  \caption{
The evolution of the fields in the last stage of inflation as a function of
e-folds $N$ for the model defined by the parameter set (\ref{par2pos}).
}
\label{2pososcfig}
\end{figure}

The scalar potential is given by
\begin{eqnarray}
V=
&&\hspace{-18pt}
\frac{e^{{\phi}^{2}}}{24t^3} \Big\{ 3 \left[  \tilde{A}\phi
+2(\phi-\phi_0){\mu}^{2} \right] ^{2}+ 2CD\left( 3{\phi}^{2}+ 4cdt^2+6
ct+6dt  \right)   {e^{- \left( c+d \right)t }}
\nonumber\\[4pt]
&&\hspace{-18pt}
+
6 C  \left[  \tilde{A} \left( {\phi}
^{2}+2ct\right)+2{\mu}^{2}\phi(\phi-\phi_0)
 \right] {e^{-ct}}+
6 D  \left[  \tilde{A} \left( {\phi}
^{2}+2dt\right)+2{\mu}^{2}\phi(\phi-\phi_0)
 \right] {e^{-dt}}
\nonumber\\[4pt]
&&\hspace{-18pt}
+{C}^{2} \left(3{\phi}^{2}+ 4{c}^{2}{t}^{2}+12ct
 \right) {e^{-2ct}}+
{D}^{2}\left( 3{\phi}^{2}+4{d}^{2}{t}^{2}+12dt
 \right) {e^{-2dt}} \Big\}
\,,
 \label{2pospot}
\end{eqnarray}
where $\tilde{A}\equiv A+\mu^2(\phi-\phi_0)^2$ and
$\phi_0\equiv {\rm Re}\Phi_0$. The imaginary part of $\Phi_0$
was chosen to vanish in order to have a SUSY (near) Minkowski minimum
at the real value of $\Phi$. During the whole period of inflation the
axions are fixed at zero so in the above formula we set $\tau=\theta=0$.
We illustrate the main features of the model by considering an example
with the following set of parameters:
\begin{equation}
\begin{tabular}{llll}
$A=-5.6547\cdot10^{-5} , \qquad$ &
$C=-2.444\cdot10^{-13} \,, \qquad$ &
$D=5\cdot10^{-8} \,, $ &
\\[4pt]
$\mu^2=4.8\cdot10^{-5} \, , \qquad$ &
$c=-\frac{2\pi}{40} \, , \qquad$ &
$d=-\frac{2\pi}{100} \, , \qquad$ &
$\phi_0=1 \, .$
\end{tabular}
\label{par2pos}
\end{equation}
The parameter $A$ is chosen in order to satisfy:
\begin{equation}
\label{Amink}
A=-C\left|\frac{cC}{dD}\right|^{\frac{c}{d-c}}
-D\left|\frac{cC}{dD}\right|^{\frac{d}{d-c}}
,
\end{equation}
which is the condition for the existence of a SUSY Minkowski minimum.
This minimum is situated at:
\begin{equation}
t_{\rm Mink}=\frac{1}{c-d}\ln\left|{\frac{cC}{dD}}\right|,
\qquad
\phi_{\rm Mink}=\phi_0
.
\end{equation}
With the parameters (\ref{par2pos}) this corresponds to
$t_{\rm Mink}\approx120.03$ and $\phi_{\rm Mink}=1$. The mass matrix at
any SUSY Minkowski minimum is always positive definite \cite{blanco}.
It was also shown in \cite{blanco} that the off-diagonal entries of
the mass matrix are proportional to the value of the gravitino mass.
Therefore, the mixing between the fields at the SUSY (near) Minkowski
minimum is strongly suppressed. In
the vacuum,
the mass of the volume
modulus, $m_t\approx m_{\tau}\approx10^{11} {\rm GeV}$, is much smaller
than that of the matter field,
$m_{\phi}\approx m_{\theta}\approx10^{13} {\rm GeV}$. However,
this hierarchy is not preserved along the inflationary trajectory
because SUSY is broken in the inflationary region.
The inflationary inflection point
is situated at $t_{\rm inf}\approx98.96$ and $\phi_{\rm inf}\approx-0.3476$.
The vanishing eigenvalue of the $\eta$-matrix corresponds to the direction
$(t, \phi)=(0.3612, -0.9325)$ so the inflaton is a mixture of the volume
modulus and the matter field with a domination of the
latter\footnote{
Looking at the inflationary trajectory in figure \ref{2pospotfig}
one could naively expect the inflaton is dominated by the volume modulus.
However, this is not really the case. One should take into account
that $t$ is a non-canonically normalized field.
Moreover, in terms of the canonically normalized fields the range of
the axis along the $\phi$ direction is much larger than the one
along the $t$ direction.}.
Other 3 eigenvalues of the $\eta$-matrix at the inflection point are
positive: $\eta_2\approx0.62$, $\eta_3\approx26.3$, $\eta_4\approx31.1$.
The stability of the directions orthogonal to the inflaton are due to
the fact that the SUSY breaking at the inflection point is dominated by
the matter field: $\Theta_{\Phi}^2|_{\rm inf}\approx0.85$.
The sectional curvature along the SUSY breaking direction is
$R(f^i)\approx0.0142$ and $\widehat{G}^2|_{\rm inf}\approx8.7$ so the
necessary condition $\ref{etacon}$ for successful inflation is satisfied.
The slow-roll parameter $\epsilon\approx10^{-9}$ at the inflection point
is small enough to allow for long lasting inflation. The numerical
integration of the equations of motion (\ref{eqmot}) shows that
inflation, starting with the fields initially set at rest at the
inflection point, ends after about $187$ e-folds. It can be seen
from figure \ref{2pososcfig} that after inflation there are very rapid
oscillations in the $t$ direction. During these oscillations and for
some time after they are damped, the matter field $\phi$ evolves
relatively slowly towards the SUSY (near) Minkowski minimum located
at $\phi_{\rm Mink}=1$ where it starts to oscillate with the initial
amplitude of about 0.1. During the matter field oscillations the
volume modulus is practically at rest at $t=t_{\rm Mink}$.


The CMB signatures of this model are similar to other models of
inflection point inflation. The parameter set (\ref{par2pos}) satisfies
COBE normalization (\ref{cobe}) for the amplitude of density
perturbations. The spectral index equals
$n_s\approx0.94$ and agrees with the formula (\ref{fns}). The tensor
to scalar perturbations ratio and the running of the spectral index are again
negligible from the observational point of view.

Let us now discuss requirements for successful inflation in the
considered model. Similarly to the original model (without the matter
field but with the string corrections) proposed in \cite{bo2},
one needs to fine-tune one parameter in the superpotential to assure
flatness of the inflection point. In the present example we choose
for that purpose the parameter $C$ which should be fine-tuned at
the level $\Delta\approx10^{-4}$ to allow for inflation longer than
$60$ e-folds. So, the fine-tuning is weaker than in the original model
without the matter field in which the parameter $C$ must be adjusted
with the precision at the level of $10^{-5}$ \cite{bo2}.

\begin{figure}[t]
\centering
  \includegraphics[width=8cm,angle=0]{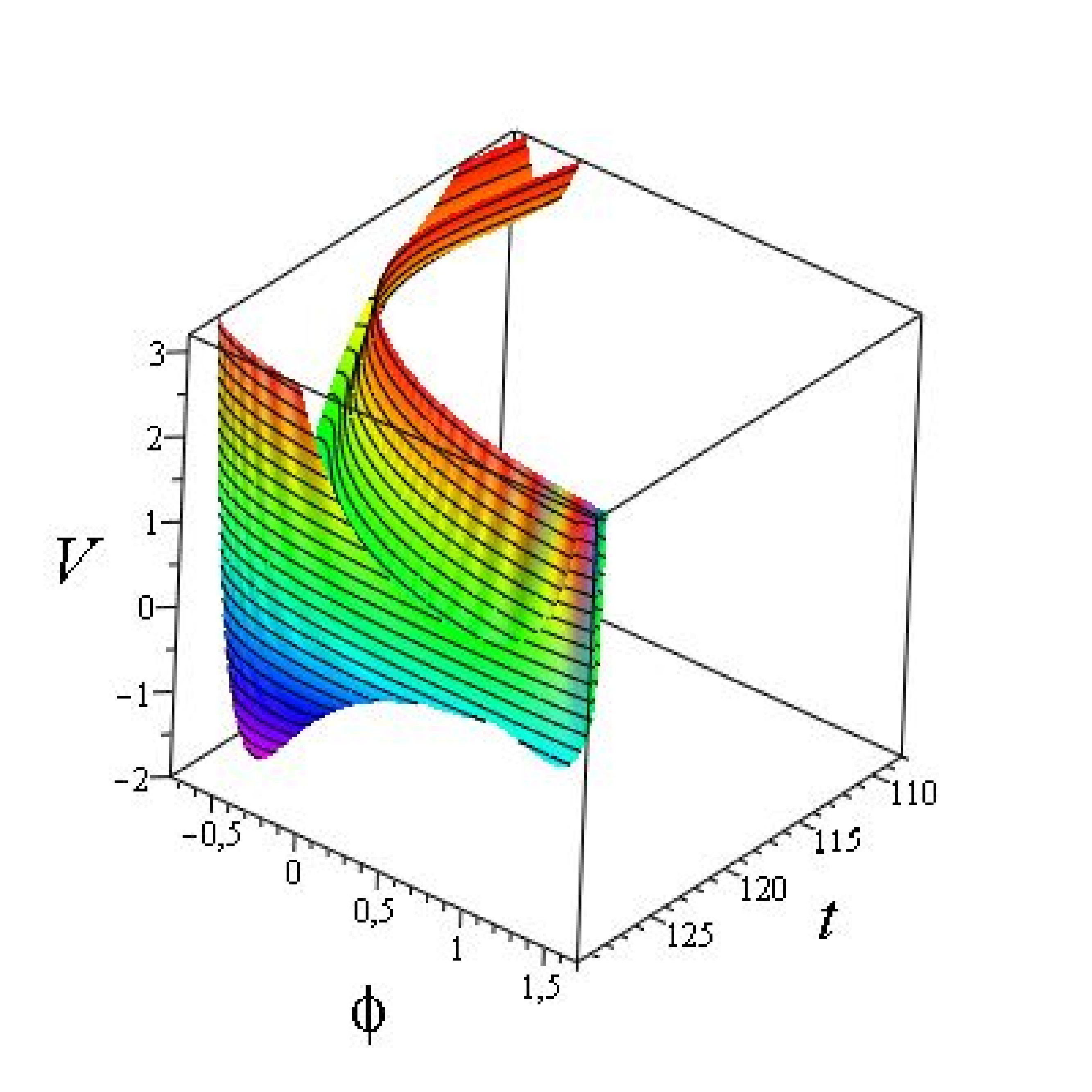}
  \caption{The potential (\ref{2pospot}) multiplied by $10^{15}$ with
$\phi_0=1.5$. The parameter $D$ is tuned to assure the existence of
a very flat inflection point, the parameter $A$ is chosen such that
the potential has a SUSY Minkowski minimum. The remaining
parameters are chosen as in (\ref{par2pos}).}
\label{2posfig_AdS}
\end{figure}

Beside the fine-tuning of the parameter $C$ making potential flat,
there is also a constraint on the parameter $\phi_0$ which defines
the position of the SUSY (near) Minkowski minimum.
We found that inflation ending in such a minimum can be realized
only for a rather small range of values of $\phi_0$ around 1.
The reason is as follows:
The value of $|\phi|$ at a typical stationary point (e.g.\
our inflationary inflection point) is smaller,
or even much smaller, than unity because of scaling of the
potential (\ref{2pospot}) by the factor $e^{\phi^2}$ which originates
from the K\"ahler potential (\ref{kahlercan}).
On the other hand, the position of a SUSY Minkowski minimum is
determined solely by the structure
of the superpotential, so it may occur at arbitrary value of $\phi$.
The potential grows exponentially with $\phi^2$ so
if a SUSY Minkowski minimum occurs at $|\phi_0|\gtrsim1$ then there
exists a barrier for $|\phi|<|\phi_0|$ which separates this SUSY
Minkowski minimum from the inflationary inflection point
located at small $\phi$. In such a case
the inflaton evolves towards a
minimum\footnote{
The vacuum in which inflaton ends its evolution for $|\phi_0|\gtrsim1$ is typically of AdS type if the parameters satisfy the condition (\ref{Amink}) for the existence of a SUSY Minkowski minimum. When one relaxes the condition (\ref{Amink}) the cosmological constant at this vacuum can be canceled with an appropriate tuning of the parameter $A$.  However, even though this vacuum can be Minkowski, the gravitino mass there is of the order of the Hubble scale, which makes that scenario less interesting.
}
which is situated between the inflationary inflection point and the
SUSY Minkowski minimum. The structure of the corresponding potential
is shown in figure \ref{2posfig_AdS}. The value of $|\phi_0|$
must not be substantially bigger than unity in order to avoid
a barrier between the vacuum and the inflection point.
On the other hand, $|\phi_0|$ can not be very small because in such
a case the contribution of the matter field to SUSY breaking at
the inflationary inflection point would not be enough to overcome
the constraint (\ref{thetaTcon}). It can be seen
from table \ref{tab_2pos_phi0} that for too small $|\phi_0|$ the
potential at the inflection point is unstable in the direction being
the mixture of the axions.
In the presented example the axionic direction
is stable at the inflationary inflection point and there is no
problem of the barrier separating that point from the vacuum for
$0.9\lesssim|\phi_0|\lesssim1.02$. So, the
position of the SUSY Minkowski minimum should be chosen with the
precision at the level of 1 part in 10.

\begin{table}[t]
\centering
\begin{tabular}{|c|c|c|c|c|}
\hline	
$\phi_0$ & $\eta_2$ & $\frac{m_2}{H}$ & $\Theta^2_T$ &  $\widehat{G}^2$
\\ \hline
$1.02$ & $0.72$ & $1.5$  & $0.14$  & $8.1$
\\ \hline
$1$ & $0.62$ & $1.4$  & $0.146$  & $8.7$
\\ \hline
$0.95$ & $0.33$ & $1$  & $0.164$  & $10.9$
\\ \hline
$0.9$ & $0.005$ & $0.1$  & $0.187$  & $14.6$
\\ \hline
$0.85$ & $-0.39$ & $-$  & $0.216$  & $21.8$
\\ \hline
\end{tabular}
\caption{
The dependence of the second smallest eigenvalue of the $\eta$-matrix,
$\eta_2$, the ratio $\frac{m_2}{H}$, $\Theta^2_T$ and $\widehat{G}^2$
on the position of a SUSY Minkowski minimum $\phi_0$. The parameter $C$
is always tuned in order to have a very flat inflection point required
for inflation. Other parameters are chosen as in (\ref{par2pos}) except
$A$ which is always chosen as to satisfy the condition (\ref{Amink})
for the existence of a SUSY Minkowski minimum. }
\label{tab_2pos_phi0}
\end{table}
\begin{table}[t]
\centering
\begin{tabular}{|c|c|c|c|c|}
\hline	
$d$ & $\eta_2$ & $\frac{m_2}{H}$ & $\Theta^2_T$ &  $\widehat{G}^2$
\\ \hline
$-\frac{2\pi}{98}$ & $0.41$ & $1.1$  & $0.153$  & $10.5$
\\ \hline
$-\frac{2\pi}{95}$ & $0.28$ & $0.9$  & $0.182$  & $12.1$
\\ \hline
$-\frac{2\pi}{90}$ & $0.2$ & $0.8$  & $0.203$  & $13.3$
\\ \hline
$-\frac{2\pi}{70}$ & $0.23$ & $0.8$  & $0.204$  & $12.9$
\\ \hline
$-\frac{2\pi}{50}$ & $0.49$ & $1.2$  & $0.165$  & $9.8$
\\ \hline
$-\frac{2\pi}{40}$ & $0.72$ & $1.5$  & $0.14$  & $8.1$
\\ \hline
$-\frac{2\pi}{30}$ & $1.08$ & $1.8$  & $0.114$  & $6.6$
\\ \hline
\end{tabular}
\hspace{10mm}
\begin{tabular}{|c|c|c|c|c|}
\hline	
$\mu^2$ & $\eta_2$ & $\frac{m_2}{H}$ & $\Theta^2_T$ &  $\widehat{G}^2$
\\ \hline
$10^{-1}$ & $1.83$ & $2.3$  & $0.043$  & $5.2$
\\ \hline
$5\cdot10^{-3}$ & $1.17$ & $1.9$  & $0.079$  & $6.4$
\\ \hline
$10^{-3}$ & $0.8$ & $1.5$  & $0.118$  & $7.7$
\\ \hline
$4.8\cdot10^{-4}$ & $0.62$ & $1.4$  & $0.146$  & $8.7$
\\ \hline
$10^{-4}$ & $0.18$ & $0.7$  & $0.243$  & $13.3$
\\ \hline
$5\cdot10^{-5}$ & $-0.031$ & $-$  & $0.314$  & $18.3$
\\ \hline
$10^{-5}$ & $-0.59$ & $-$  & $0.605$  & $180$
\\ \hline
\end{tabular}
\caption{
The dependence of the same quantities as in table \ref{tab_2pos_phi0}
on the parameter $c$ (left) and $\mu^2$ (right). The position of the SUSY
Minkowski minimum is fixed at $\phi_0=1$. Other parameters are chosen
as in (\ref{par2pos}) except $C$, which is always tuned to have a very
flat inflection point appropriate for inflation, and $A$,
which controls the existence of a SUSY Minkowski minimum. }
\label{tab_2pos}
\end{table}

A detailed analysis of the potential (\ref{2pospot}) shows that the
axionic direction becomes more stable when the ratio $|c/d|$ deviates
more from unity. This can be seen in table  \ref{tab_2pos} (left).
The axionic direction is also more stable for the value of $c$ very
close to that of $d$. However, smaller differences between $c$ and $d$ seems
to be less natural because the fine-tuning of the parameter $C$ required
to obtain more than $60$ e-folds of inflation is  stronger in such cases.
Finally, only for large enough values of $\mu^2$ the axionic direction
can be stable, as can be seen from table \ref{tab_2pos} (right).
This is rather intuitive because $\mu^2$ controls the magnitude of
the matter field $F$-term so for small values of $\mu^2$ the contribution
of the matter field to the SUSY breaking is to small to satisfy the
necessary condition (\ref{thetaTcon}).

We should also add that for many sets of the parameters which allow for
inflation, the second smallest eigenvalue of the $\eta$-matrix
corresponding to the axionic direction is much smaller than unity
and in consequence the corresponding mass eigenvalue is smaller
than the Hubble scale during inflation. In such cases the quantum
fluctuations of this axionic direction can give rise to
isocurvature perturbations which can significantly affect the power
spectrum of the CMB \cite{isocurvature}. One can enlarge the
eigenvalue of the $\eta$-matrix corresponding to the axionic
direction by increasing $\mu^2$, increasing the difference between
$c$ and $d$ or by taking $|\phi_0|$ close to the maximal value
for which inflation ends in a SUSY (near) Minkowski minimum.
This can be seen from tables \ref{tab_2pos_phi0} and \ref{tab_2pos}.
Significant production of the isocurvature perturbations can be
avoided also when $c$ and $d$ have very similar values, as can be
seen from the left panel of table \ref{tab_2pos}.
In the presented example the mass of the axionic direction
$m_2\approx9\cdot10^{10} {\rm GeV}$ during inflation is larger
than the Hubble scale $H\approx7\cdot10^{10} {\rm GeV}$ so the
isocurvature perturbations are not significantly produced.

\subsection{Matter field inflation}

In the model presented in the previous subsection the matter field
plays an important role in the inflationary dynamics and one cannot
realize ''pure'' volume modulus inflation ending in a SUSY (near)
Minkowski minimum. Therefore, since the matter field has to contribute
to the inflationary sector anyway it seems sensible to investigate
models in which SUSY breaking at the inflationary inflection point
and the whole inflationary dynamics is dominated by the matter field.

The main problem of the model considered in the previous subsection
is the existence of the barrier at $|\phi_0|\gtrsim1$ which separates
the inflationary inflection point from a SUSY (near) Minkowski minimum.
From the viewpoint of inflation driven by the matter field this feature
of the potential is an advantage because for appropriate choice of
$\phi_0$ one can obtain a very flat inflection point.
The inflationary inflection point appears for a value of $|\phi_0|$
slightly smaller than the one for which a barrier arises.
Interestingly, that kind of model can be constructed for any combination
of signs of the exponents $c$ and $d$ in (\ref{Wrace}). Moreover,
independently from the signs of the exponents inflation can be realized
in the vicinity of the inflection point situated at
$|\phi_{\rm inf}|\ll1$ for $|\phi_0|\approx 1$. The similarities
between models with different sign assignments follows from the
fact that inflection point inflation can be realized in a model
with the superpotential (\ref{Wmatter2}) without the inclusion of
the volume modulus sector. That kind of model was proposed long
time ago \cite{Holman}. The scalar potential for the matter field
in that model is given by:
\begin{equation}
\label{potmatter}
V(\phi)={\mu}^{4}{e^{{\phi}^{2}}}\left(\phi-\phi_0\right)^{2}\left[  \left( {\phi}^{2}-3 \right) \left(\phi-\phi_0\right)^{2}+4\phi\left(\phi-\phi_0\right) +4\right]
\end{equation}
For $\phi_0=1$ the first and the second derivative of the above
potential at $\phi=0$ vanish. This implies that inflation
ending in a SUSY Minkowski minimum at $\phi\approx1$ can be
realized in the vicinity of an inflection point located
at very small values of $\phi$.

\begin{figure}[t]
\centering
  \includegraphics[width=8cm,angle=0]{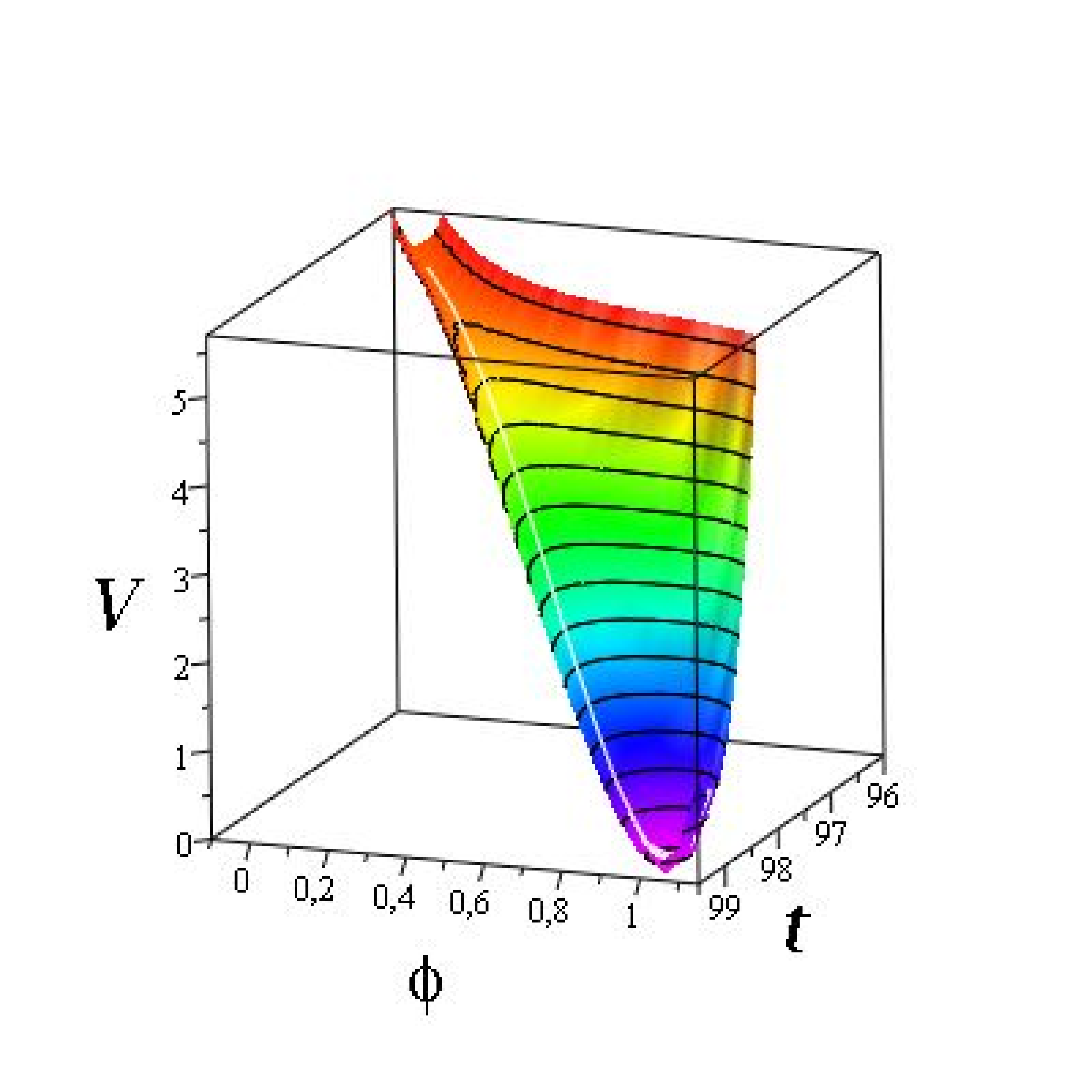}
  \caption{The potential (\ref{2pospot}) multiplied by $10^{15}$ for the model defined by the parameter set (\ref{parmatterKL}). The white curve represents the inflationary trajectory from the inflection point to the SUSY (near) Minkowski minimum.}
\label{KLmatterfig}
\end{figure}

After coupling the model to the volume modulus sector with the
superpotential (\ref{Wrace}) the scalar potential is given by
(\ref{2pospot}). We found that the inclusion of the volume modulus
sector only mildly affects the original model (\ref{potmatter}).
In order to illustrate this we present an example of a model
coupled to the KL model with the following set of parameters:
\begin{equation}
\begin{tabular}{llll}
$A=-1.9531\cdot10^{-4} , \qquad$ &
$C=-1.8 \,, \qquad$ &
$D=1 \,,$ &
\\[4pt]
$\mu^2=2\cdot10^{-4} \, , \qquad$ &
$c=\frac{2\pi}{90} \, , \qquad$ &
$d=\frac{2\pi}{100} \, , \qquad$ &
$\phi_0=1.04752 \, .$
\end{tabular}
\label{parmatterKL}
\end{equation}
The potential in ($t$, $\phi$) subspace is shown in figure
\ref{KLmatterfig}. The parameter $A$ is chosen as to guarantee
the existence of the SUSY (near) Minkowski minimum which occurs
at $t_{\rm Mink}\approx99.29$ and $\phi_{\rm Mink}\approx1.04572$.
Inflation takes place in the vicinity of the inflection point
which is situated at $t_{\rm inf}\approx96.55$ and
$\phi_{\rm inf}\approx0.0617$. The SUSY breaking at this inflection
point is strongly dominated by the matter field:
$\Theta_{\Phi}^2|_{\rm inf}\approx0.9981$. The sectional curvature
along the SUSY breaking direction is very small:
$R(f^i)\approx2\cdot10^{-6}$.
So, the necessary condition (\ref{etacon}) for the stability of
the directions orthogonal to the inflaton is easily satisfied.
The $\eta$-matrix at the inflection point has 3 positive eigenvalues:
 $\eta_2\approx3.88$, $\eta_3\approx6375$, $\eta_4\approx6387$.
The remaining, vanishing eigenvalue corresponds to the direction
($t$, $\phi$)=(0.0426, 0.9991), so the matter field is the main
component of the inflaton. The second slow-roll parameter, $\epsilon$,
has at the inflection point a very small value of the order of
$2\cdot10^{-9}$. Starting with the fields set initially at rest
at the inflection point, one obtains about 102 e-folds of inflation
which ends in a SUSY (near) Minkowski minimum.
The inflationary trajectory is shown in figure \ref{KLmatterfig}.

The spectral index depends on the number of e-folds obtained during
inflation and is well approximated by the formula (\ref{fns}).
In the example with the parameters (\ref{parmatterKL}) we found
$n_s\approx0.94$ and unmeasurably small values of the tensor to
scalar perturbation amplitudes ratio and the running of the
spectral index. This model does not predict generation of the
isocurvature perturbations since the second smallest eigenvalue
of the $\eta$-matrix, $\eta_2$, during inflation is larger than unity.

The model considered in this subsection requires the fine-tuning
of the position of the SUSY Minkowski minimum $\phi_0$ which has
to be chosen with the precision $\Delta\approx10^{-5}$ in order
to get more than $60$ e-folds of inflation. In addition, the parameter
$\mu^2$ cannot be larger than some critical value (depending on the
values of other parameters) above which inflaton overshoots the SUSY
Minkowski minimum and runs away to the region of infinite volume.
This constraint corresponds to an upper bound on the mass of the matter
field (controlled by $\mu^2$). The reason for the existence of such a
bound is as follows: The inflationary inflection point occurs at smaller
volume than the SUSY (near) Minkowski minimum, i.e.\
$t_{\rm inf}<t_{\rm Mink}$. If the volume modulus is much heavier than
the matter field, the value of $t$ does not differ significantly between
the inflationary inflection point and the SUSY Minkowski minimum.
The distance in the $t$ direction between the inflection point and the
Minkowski minimum, $t_{\rm Mink}-t_{\rm inf}$, grows when the mass of the
volume modulus decreases (as compared to the mass of the matter field).
Also the distance in the $t$ direction between the inflection point and
the barrier, $t_{\rm bar}-t_{\rm inf}$, grows in such a case.
As a result, the difference between the values of the potential
at the inflection point and the barrier becomes bigger
(due to the prefactor $t^{-3}$ present in the potential (\ref{2pospot})).
In consequence, for a too light volume modulus the height of the barrier
becomes to small and the inflaton overshoots the SUSY Minkowski minimum.
Numerical calculations show that the mass of the matter field
should be at least about two times smaller than the mass of the
volume modulus in order to avoid overshooting of the SUSY (near)
Minkowski minimum. In the example used in this subsection the mass
of the matter field
$m_{\phi}\approx m_{\theta}\approx6\cdot10^{11} {\rm GeV}$
is much smaller than the mass of the volume modulus
$m_t\approx m_{\tau}\approx2\cdot10^{13} {\rm GeV}$ so
the overshooting problem is absent.

The presented example assumes KL model of moduli stabilization.
However, we would like to emphasize that very similar models can
be constructed also for positive exponents in non-perturbative
terms in the racetrack superpotential (\ref{Wrace}). Moreover,
if at least one exponent is positive then inflation may end in a
SUSY (near) Minkowski minimum for any value of the parameter $\mu^2$.

Since in this section we focus on models with a very light gravitino,
in our numerical example the parameter $A$ is tuned in order to have
a SUSY (near) Minkowski minimum. However,
inflation can be realized also without this tuning i.e.\ with the
standard non-SUSY uplifting of the form (\ref{uplift}) generating
a large gravitino mass of the order of the Hubble scale.
Moreover, if one gives up a light gravitino, one does not need to
introduce two non-perturbative terms in the superpotential -
one gaugino condensate of the standard KKLT model \cite{kklt}
is enough. Therefore, we conclude that the inflection point
model with the superpotential (\ref{Wmatter2}) can be easily
coupled to the volume modulus sector. This result is not very
sensitive to the choice of the moduli stabilization mechanism.

\subsubsection{Modular weights}

We now study the situation in which the matter field has non-vanishing
modular weight. In such a case the K\"ahler potential is given by:
\begin{equation}
\label{kahlerweight}
K=-3\ln(T+\ov{T})+\frac{\Phi\ov{\Phi}}{(T+\ov{T})^n} \,,
\end{equation}
where $n$ is a positive integer. Notice that the K\"ahler potential
is no longer separable and generates a non-diagonal K\"ahler metric.
In consequence, the curvature tensor of the corresponding K\"ahler
manifold has non-vanishing mixed components. From the explicit formulae
for all components (collected in appendix B) one can infer
some useful information. Namely, the components of the curvature tensor
with less than two $T$-indices vanish for any $n\neq0$. Since we focus
on a model in which the Goldstino direction $f^i$ at the inflationary
point is dominated by the $\Phi$-direction, we expect that the main
contribution to $R(f^i)$ comes from terms proportional to the
components of the curvature with zero and one $T$-index. Therefore,
$R(f^i)$ should remain very small also for $n\neq0$ in favor of
slow-roll inflation.

We confirmed our expectations by the numerical analysis. Using KL model
for moduli stabilization, we found that one can construct very flat
inflection point also for a non-vanishing modular weight. In such a case,
the inflationary scenario is very similar to the one with a canonically
normalized matter field. The position of the SUSY (near) Minkowski
minimum, as well as the position of the inflationary inflection point,
in terms of canonically normalized matter field
$\Phi_{\rm can}\equiv\frac{\Phi}{(T+\ov{T})^{n/2}}$, is typically of the
same order for any $n$.

The following comment on the $n=1$ case is in order.
The corresponding K\"ahler potential (\ref{kahlerweight}) can be obtained
from the expansion of the no-scale K\"ahler potential
(\ref{kahlernoscale}) for $|\Phi|^2/(T+\ov{T})\ll1$.
In the $n=1$ model inflation takes place for
$|\Phi|^2/(T+\ov{T})\ll1$. So, one could naively expect that
the model with the superpotential (\ref{kahlernoscale})
should have properties similar to the model with the superpotential
(\ref{kahlerweight}) for $n=1$. However, this is not the case.
The overall contribution of the subleading terms in the expansion
of (\ref{kahlernoscale}) is important and results in the constant
curvature $R(f^i)=2/3$ of the corresponding K\"ahler manifold so
the necessary condition (\ref{etacon}) is violated \cite{Grs2}.

\section{Conclusions}
\label{conclsec}

Our goal has been to construct supersymmetric models of inflation
involving the volume modulus and a matter field. The primary
purpose of including the matter field was to use its F-term to
uplift the energy of the vacuum to a small positive value. In most
of the models proposed so far, the uplifting was achieved by adding
some terms which break SUSY explicitly (e.g.\ terms associated with
$\overline{D3}$-branes). The uplifting can be realized in a fully
supersymmetric way if the matter field is used. There is also
another very important role of the matter field. Contributing to
SUSY breaking, it can stabilize the directions perpendicular
to the inflaton, some of which are usually tachyonic in supersymmetric
models involving only moduli.

We concentrated on models in which the superpotential and the
K\"ahler potential can be written as sums of terms depending
on just one field, the volume modulus $T$ or the matter field $\Phi$.
For the volume modulus contributions we used the standard
no-scale K\"ahler potential and the superpotential of the
racetrack form. We investigated several models with
different K\"ahler potentials and superpotentials for the
matter field looking for such values of the parameters
which can support a period of long enough slow roll inflation.
Two scenarios have been considered. In one of them inflation
takes place when the fields have values close to
a saddle point of the potential. In the second scenario
they are close to an inflection point.

We found that it is relatively easy to construct models
of inflation if one does not insist on having a low scale of
SUSY breaking. The superpotential for the matter field
may have a simple form linear in $\Phi$.
We considered two K\"ahler potentials for the matter field:
the canonical one and the canonical one with a correction
proportional to $(\Phi\bar\Phi)^2$. In these two cases
the matter sector is the same as in the Polonyi model or in
the quantum corrected O'Raifeartaigh model, respectively.
For both cases we presented examples of saddle point and
inflection point inflation.
The models of inflection point inflation can be treated as
effectively involving only two real fields because the
imaginary components of the modulus and the matter field can
be set to zero during and after inflation. All four real fields
are to be taken into account in the case of
the saddle point inflation models.

The main difference between models with different matter
sectors is the role of the matter field in the inflationary
dynamics. In models with the Polonyi uplifting,
the inflaton is a mixture of the modulus and the matter field
with a bigger contribution from the latter. For the case of
the O'Raifeartaigh uplifting, the matter field is decoupled
from the inflationary dynamics and the inflaton can be almost
a pure modulus field.

In all models SUSY breaking at the inflationary region
is dominated by the matter field but the strength of
this dominance changes
from case to case. It is much stronger for the inflection
point models as compared to the saddle point ones. Within
each type of models (inflection point or saddle point)
the dominance of matter field SUSY breaking is stronger
for the Polonyi uplifting. The reason is that the matter
field submanifold has vanishing curvature for the Polonyi
model while it is negative for the O'Raifeartaigh one.
In the inflection point inflation models, the volume modulus
contribution to SUSY breaking is at the level of one (a few)
percent in the case of the Polonyi (O'Raifeartaigh) uplifting.
In the saddle point models this contribution is
typically bigger by one order of magnitude.
For the saddle point model
with the O'uplifting the contribution to SUSY breaking
from the modulus may be even comparable to that of the
matter field.

In the inflationary region there are three scalars
with positive masses squared and one tachyonic state.
In all the model investigated in this paper the inflaton
and the lightest non-tachyonic state are dominated by two
components of the same complex field (the modulus or the
matter field). We found that the value of the inflationary
Hubble constant is smaller than the masses of all scalar fields
(other than the inflaton), so the isocurvature perturbations
are suppressed. This suppression is stronger in models
with O'uplifting.

The fine tuning of the superpotential parameters necessary
to obtain a long enough period of inflation depends on the model.
In the case of the saddle point models this fine tuning
varies between a few times $10^{-4}$ (O'uplifting)
and a few times $10^{-3}$ (Polonyi uplifting),
so it is somewhat weaker than for analogous models with
non-SUSY uplifting (typically at the level of $10^{-4}$).
From the point of view of fine tuning the Polonyi
uplifting seems to be better than the O'uplifting.
This feature is more pronounce in the inflection point scenario.
O'uplifting requires the constant term in the superpotential
to be adjusted with the precision at the level $10^{-9}$ which
is worse than $10^{-8}$ typical for inflection point models
with non-SUSY uplifting. Using the Polonyi uplifting allows
to decrease necessary fine tuning by four or even five orders of
magnitude to the level $10^{-4}-10^{-3}$.

Another goal of the paper has been to construct models of inflation
ending in  SUSY (near) Minkowski minimum. The main motivation for
building such models is that they predict a low scale of SUSY breaking
in contrast to the models with uplifting summarized above. This kind
of models involving only the volume modulus have been constructed
before \cite{bo,bo2}. However, in those models the string corrections
to the K\"ahler potential have to be taken into account in order to
decrease the curvature of the K\"ahler manifold and fulfill the
necessary condition for slow-roll inflation. In the present paper we
addressed the question whether the contribution to SUSY breaking from
the matter field could decrease the curvature of the K\"ahler manifold
in such a way that inflation could be realized without inclusion of
any string corrections to the K\"ahler potential. In particular, we
investigated a model with the racetrack superpotential with two
positive exponents. We assumed quadratic superpotential for the
matter field which is the minimal choice allowing for the existence
of SUSY Minkowski minimum. We found that inflation in such model can
be realized if one parameter from the modulus sector is fine-tuned
with the precision at the level of $10^{-4}$. However, the matter
field dominates the inflaton field so ''pure'' volume modulus inflation
cannot be achieved. Moreover, the parameter space of the matter field
sector allowing for inflation is significantly constrained. In addition,
the demand for the absence of significant isocurvature perturbations puts some extra constraints on the parameter space of the model because for many parameter sets the mass of the lightest non-tachyonic state during inflation is smaller than the Hubble scale.

We found that it is much easier to construct models of inflation
ending in a SUSY (near) Minkowski minimum if the inflaton is totally
dominated by the matter field. In such a case one only needs to
fine-tune the vev of the matter field. Inflation can be realized
for any combination of signs of the exponents in the racetrack
superpotential including two negative ones which correspond to
the KL model of moduli stabilization. However, in the case of KL model
the volume modulus cannot be lighter than the matter field. Otherwise
the inflaton would overshoot the SUSY Minkowski minimum and evolve
towards the region of infinite volume. It is not necessary to restrict
to the canonical K\"ahler potential for the matter fields because
similar models can be also constructed for non-vanishing modular
weight of the matter field. In the models with inflation dominated
by the matter field masses of all non-tachyonic states are much bigger
than the Hubble scale during inflation so the generation of the
isocurvature perturbations is strongly suppressed.

The spectral index in all models discussed in this paper depends on
the specific choice of parameters but cannot take arbitrary values.
Its allowed values depend on whether inflation takes place in
the vicinity of an inflection point or a saddle point.
In the former case spectral index is bounded from below $n_s\gtrsim0.93$,
while in the latter case it is bounded from above $n_s\lesssim0.95$.
In any case the parameters can be chosen as to fit the WMAP5 data.
The tensor to scalar perturbation ratio and the running of the
spectral index predicted by all models under consideration are
unmeasurably small.

In summary, our investigation shows that in models admitting a low
scale of SUSY breaking it seems to be much more natural to consider
the matter field, rather than the volume modulus, as the inflaton.
On the other hand, if one does not insist on a low scale of SUSY
breaking, the volume modulus is also a very good candidate for the inflaton.
In particular, the volume modulus can dominate inflation in the racetrack
models uplifted by the matter field $F$-term even though the matter
field dominates SUSY breaking. Since the matter field sectors that we
used in this work are described by superpotentials and K\"ahler
potentials of very simple forms, we believe that in more realistic
constructions, rigorously derived from the string theory, the $F$-term
uplifting of racetrack inflation would be also accomplished.
We hope that our analysis will be a useful guideline for the future work.

\section*{Acknowledgments}

This work was partially supported by the EC 6th
Framework Projects
MTKD-CT-2005-029466 ``Particle Physics and Cosmology:
the Interface'' and MRTN-CT-2006-035863
``The Origin of Our Universe: Seeking Links between
Fundamental Physics and Cosmology''.
M.O.\ would like to thank for the hospitality experienced
at Max Planck Institute for Physics in Munich, where
part of this work has been prepared.

\appendix
\section{Equations of motion and CMB observables}
\renewcommand{\theequation}{A.\arabic{equation}}
\setcounter{equation}{0}

In order to study the evolution of the fields during inflation one has
to solve the appropriate equations of motion. The equation for the real
fields are needed while the metric $G_{i\ov{j}}$ computed using the
K\"ahler potential defines the kinetic terms for the complex fields.
The relation between the metric for the complex fields $\Phi^i$ and
the metric $g_{ij}$ for the real fields $\phi^i$ can be read off from
the scalar part of the 4D SUGRA action:
\begin{equation}
S_{\rm scalar}=-\int{\rm d}^4x\sqrt{-g}\left[G_{i\ov{j}}\pa_{\mu}\Phi^i\pa^{\mu}\ov{\Phi^j}+V\right]=-\int{\rm d}^4x\sqrt{-g}\left[\frac{1}{2}g_{ij}\pa_{\mu}\phi^i\pa^{\mu}\phi^j+V\right] \,.
\end{equation}
The equations of motion resulting from the above action read:
\begin{eqnarray}
        \ddot{\phi^i}+3H\dot{\phi^i}+\Gamma_{jk}^i\dot{\phi^j}\dot{\phi^k}
+g^{ij}\frac{\pa V}{\pa \phi^j}=0\,,
\label{EoM_phi}
\\
H^2=\left(\frac{\dot{a}}{a}\right)^2
=\frac{1}{3}\left(\frac{1}{2}g_{ij}\dot{\phi^i}\dot{\phi^j}+V\right)\,,
\label{EoM_a}
\end{eqnarray}
where $a$ is the cosmic scale factor, $H$ is the Hubble parameter, dots
denote derivatives with respect to the cosmic time and $\Gamma_{jk}^i$ are
the standard Christoffel symbols given in terms of the metric $g_{ij}$ by:
\begin{equation}
\Gamma_{jk}^i=\frac{1}{2}g^{il}\left[\frac{\pa g_{lk}}{\pa\phi^j}
+\frac{\pa g_{jl}}{\pa\phi^k}-\frac{\pa g_{jk}}{\pa\phi^l}\right]
.
\end{equation}
It is convenient to study the field evolution using the number of e-folds,
$N$, related to the scale factor and the cosmic time by:
\begin{equation}
        a(t)=e^N, \hspace{2cm}\frac{d}{dt}=H\frac{d}{dN}\,.
\end{equation}
Using the above definition and eq.\ (\ref{EoM_a}), the equation of motion
(\ref{EoM_phi}) may be rewritten \cite{deCarlos} in the form:
\begin{equation}
\label{eqmot}
\phi^{i''}+\left[3-\frac{1}{2}g_{jk}\phi^{j'}\phi^{k'}\right]\left[\phi^{i'}
+g^{ij}\frac{1}{V}\frac{\pa V}{\pa \phi^j}\right]
+\Gamma_{jk}^i\phi^{j'}\phi^{k'}=0
\,,
\end{equation}
where $'$ denotes derivatives with respect to $N$.

Every inflationary model has to satisfy the COBE normalization \cite{lyth}:
\begin{equation}
\label{cobe}
 \frac{\delta\rho}{\rho}=\frac{2}{5}\sqrt{{\cal P_R}(k_0)}
\approx 2\cdot10^{-5}
\ ,
\end{equation}
where $k_0\approx7.5H_0$ is the COBE normalization scale which leaves
the horizon approximately $55$ e-folds before the end of inflation and
${\cal P_R}$ is the amplitude of the scalar perturbations given, in
the slow-roll approximation, by the following formula:
\begin{equation}
{\cal P_R}(k)=\frac{1}{24\pi^2}
\left.\left(\frac{V}{\epsilon}\right)\right|_{k=aH} \ .
\end{equation}
The r.h.s.\ of the above equation is evaluated at the time when
the scale $k$ crosses the horizon and $\epsilon$ is the generalized
slow-roll parameter:
\begin{equation}
\epsilon\equiv\frac{1}{2}g^{ij}\left(\frac{V_iV_j}{V^2}\right) \ ,
\end{equation}
where $V_i\equiv\frac{\pa V}{\pa \phi^i}$.

Very important observable is the spectral index defined as follows:
\begin{equation}
        n_s-1\equiv\frac{d\ln\mathcal{P}_{\mathcal{R}}(k)}{d\ln k}
\approx\frac{d\ln\mathcal{P}_{\mathcal{R}}(N)}{dN}\ .
\end{equation}
The last approximation comes from the fact that this quantity is
evaluated at horizon crossing $k=aH=He^N$ which implies
$d\ln k\approx dN$. The spectral index can be also expressed in terms
of slow-roll parameters as $n_s\approx1+2\eta-6\epsilon$.
The generalized slow-roll parameter $\eta$ is the
smallest eigenvalue of the $\eta$-matrix \cite{geneta}:
\begin{equation}
\eta_i^j=\frac{g^{jk}\nabla_i\nabla_kV}{V} \ ,
\end{equation}
where $\nabla_i\nabla_kV=\pa_i\pa_k V-\Gamma_{ik}^l\pa_l V$.

\section{Curvature tensor for non-diagonal K\"ahler metric}
\renewcommand{\theequation}{B.\arabic{equation}}
\setcounter{equation}{0}

In this appendix we compute a curvature tensor for a non-diagonal
K\"ahler metric defined by the K\"ahler potential (\ref{kahlerweight})
for the matter field with a non-vanishing modular weight. General
formula for the curvature tensor for the K\"ahler manifold reads:
\begin{equation}
R_{i\ov{j}p\ov{q}}=K_{i\ov{j}p\ov{q}}-K_{ip\ov{r}}G^{\ov{r}s}K_{s\ov{j}\ov{q}}
\,.
\end{equation}
The non-vanishing components of the curvature tensor derived from
the above formula using the K\"ahler potential (\ref{kahlerweight})
are given by:
\begin{eqnarray}
\label{curvtensweight}
R_{T\ov{T}T\ov{T}}={\frac {18+ 2{n}^{3}\left( 6{\mathcal K}_m +{\mathcal K}_m^{2} \right) + 2{n}^{2}\left( 3{\mathcal K}_m +{\mathcal K}_m^{2} \right) +12n{\mathcal K}_m }{ \left( T+\ov{T} \right) ^{4} \left( 3
+n{\mathcal K}_m  \right) }}
 \nn[4pt]
R_{\Phi\ov{T}T\ov{T}}=R_{T\ov{T}\Phi\ov{T}}={\frac {-6{n}^{2}\ov{\Phi} }{ \left( T+\ov{T} \right) ^{3+n} \left( 3
+n{\mathcal K}_m  \right)}}
\nn[4pt]
R_{T\ov{T}T\ov{\Phi}}=R_{T\ov{\Phi}T\ov{T}}={\frac {-6{n}^{2}\Phi }{ \left( T+\ov{T} \right) ^{3+n} \left( 3
+n{\mathcal K}_m  \right)}}
\nn[4pt]
R_{T\ov{T}\Phi\ov{\Phi}}=R_{\Phi\ov{\Phi}T\ov{T}}=R_{\Phi\ov{T}T\ov{\Phi}}=R_{T\ov{\Phi}\Phi\ov{T}}={\frac {3n }{\left( T+\ov{T} \right) ^{2+n}\left( 3
+n{\mathcal K}_m  \right)}}
\end{eqnarray}
For convenience we introduced
${\mathcal K}_m\equiv\frac{|\Phi|^2}{(T+\ov{T})^n}$. Notice that many
components of $R_{i\ov{j}p\ov{q}}$ vanish including all of those with
less than two $T$-indices.

\end{document}